\documentclass[
  journal=pasa,
  manuscript=research-paper,
  year=2025,
  volume=37,
]{cup-journal}

\usepackage{amsmath}
\usepackage[nopatch]{microtype}
\usepackage{booktabs}
\usepackage{rotating} 
\usepackage{amssymb}
\usepackage{graphicx}
\usepackage{float}
\usepackage{multirow}
\usepackage{tikz}
\usepackage{xspace}
\usepackage{commath}
\usepackage{orcidlink}
\usepackage{nicematrix}
\usepackage[dvipsnames]{xcolor}

\newcommand{\ppxf}{\texttt{\textsc{pPXF}}\xspace}
\newcommand{\rtwo}{R$_{200}$\xspace}

\newcommand{\hb}{H$\beta$\xspace}
\newcommand{\oiii}{[OIII]$\lambda$5007\xspace}
\newcommand{\nii}{[NII]$\lambda$6584\xspace}
\newcommand{\hd}{H$\delta$\xspace}
\newcommand{\lt}{$<$\xspace}
\newcommand{\gt}{$>$\xspace}

\newcommand{\lsim}{$\lesssim$\xspace}
\newcommand{\gsim}{$\gtrsim$\xspace}
\newcommand{\fraction}[3]{$#1^{+#2}_{-#3}$\%}

\newcommand{\s}{$\sim$}
\newcommand{\ha}{H$\alpha$\xspace}
\newcommand{\kms}{$\,\rm{km\,s^{-1}}$}
\newcommand{\AAs}{\AA\xspace}
\newcolumntype{?}[1]{!{\vrule width #1}}
\hypersetup{colorlinks,citecolor=blue,linkcolor=blue,urlcolor=blue}

\title{The Impact of Cluster Mergers on Galaxy Properties}

\author{\orcidlink{0000-0002-1045-2559} Oğuzhan Çakır}
\affiliation{School of Mathematical and Physical Sciences, Macquarie University, Sydney, NSW 2109, Australia}
\alsoaffiliation{Astrophysics and Space Technologies Research Centre, Macquarie University, Sydney, NSW 2109, Australia}
\alsoaffiliation{ARC Centre of Excellence for All-Sky Astrophysics in 3 Dimensions (ASTRO 3D), Australia}
\email[Oğuzhan Çakır]{oguzhan.cakir@students.mq.edu.au}

\author{\orcidlink{0000-0002-2879-1663} Matt S. Owers}
\affiliation{School of Mathematical and Physical Sciences, Macquarie University, Sydney, NSW 2109, Australia}
\alsoaffiliation{Astrophysics and Space Technologies Research Centre, Macquarie University, Sydney, NSW 2109, Australia}
\alsoaffiliation{ARC Centre of Excellence for All-Sky Astrophysics in 3 Dimensions (ASTRO 3D), Australia}

\author{\orcidlink{0009-0006-8337-8712} Lucas C. Kimmig}
\affiliation{Universit\"ats-Sternwarte, Fakult\"at f\"ur Physik, Ludwig-Maximilians-Universit\"at M\"unchen, Scheinerstr.\ 1, 81679 M\"unchen, Germany}

\author{\orcidlink{0000-0003-0297-4493} Paul Nulsen}
\affiliation{ICRAR, University of Western Australia, 35 Stirling Hwy, Crawley, WA 6009, Australia}
\alsoaffiliation{Center for Astrophysics | Harvard \& Smithsonian, 60 Garden St, Cambridge, MA 02138, USA}

\author{\orcidlink{0000-0002-5896-0034} Mina Pak}
\affiliation{School of Mathematical and Physical Sciences, Macquarie University, Sydney, NSW 2109, Australia}
\alsoaffiliation{ARC Centre of Excellence for All-Sky Astrophysics in 3 Dimensions (ASTRO 3D), Australia}
\alsoaffiliation{Korea Astronomy and Space Science Institute (KASI), 776 Daeduk-daero, Yuseong-gu, Daejeon  34055, Republic of Korea}

\author{\orcidlink{0009-0009-9074-716X} Gabriella Quattropani}
\affiliation{School of Mathematical and Physical Sciences, Macquarie University, Sydney, NSW 2109, Australia}
\alsoaffiliation{Astrophysics and Space Technologies Research Centre, Macquarie University, Sydney, NSW 2109, Australia}
\alsoaffiliation{ARC Centre of Excellence for All-Sky Astrophysics in 3 Dimensions (ASTRO 3D), Australia}

\author{\orcidlink{0000-0001-5005-3125} Warrick J. Couch}
\affiliation{Centre for Astrophysics and Supercomputing, Swinburne University, Hawthorn VIC 3122, Australia}


\begin{document}

\begin{abstract}
 The impact of galaxy cluster mergers on the properties of the resident galaxies remains poorly understood. In this paper, we investigate the effects of merging environments on star formation (SF) activity in nearby clusters ($0.04<z<0.06$) from the SAMI Galaxy Survey --- A168, A2399, A3380, and EDCC 0442 --- which exhibit different dynamical activity. Using single-fibre spectroscopy from the SAMI Cluster Redshift Survey and Sloan Digital Sky Survey, we trace SF activity across the cluster sample by identifying the star-forming galaxy (SFG) population based on spectral features. We find a mild enhancement in the star-forming galaxy fraction ($f_{SFG}$) in merging clusters, although not statistically significant. The spatial and projected phase-space distributions show that SFGs in merging clusters are well-mixed with the passive population, while galaxy populations exhibit a clear segregation in the relaxed clusters. Analysis of the equivalent width of the \ha line, as a tracer of recent SF activity, does not reveal strong evidence of triggered SF activity as a function of dynamical state for both the global cluster environment and subsamples of galaxies selected near possible merger features. This suggests that the increase in $f_{SFG}$ is due to the mixing of galaxies in dynamically complex, young merging systems that are still forming, unlike their older, relaxed counterparts that have had longer to quench.
\end{abstract}

\section{Introduction} \label{sec:intro}
During the hierarchical formation of structure, galaxy clusters grow through constant infall of galaxies from surrounding filaments, the accretion of groups, and merging with other clusters \citep{Sarazin2002}. Among these accretion mechanisms, major mergers between clusters of roughly equal mass are the most energetic events (${\sim 10^{63-64}}$ erg) since the Big Bang \citep{Markevitch1998}. A significant fraction of galaxy clusters at $z < 1$ are still undergoing mergers \citep{JF1999, Mann2012}, raising the key question of how such events affect the constituent galaxies. 
\\
\\
In the local universe, we have a good understanding of the link between galaxy environments and galaxy properties such as morphology \citep{Dressler1980} and star formation (SF) activity \citep{Lewis2002, vdL2010, Barsanti2018}. Clusters are predominantly populated by red, passive early-type galaxies, whereas blue, late-type star-forming galaxies (SFGs) are more commonly found in low-density environments. A significant fraction of cluster galaxies may already be partially/fully quenched prior to their infall into cluster environments. This quenching is known as pre-processing and can contribute to observed quenched galaxy fractions out to 5\rtwo \citep{McGee2009, Haines2015, Rhee2020, Piraino-Cerda2024}. Within \rtwo, cluster-specific mechanisms can be divided into two main classes of quenching mechanisms, namely gravitational and hydrodynamical interactions.
\\
\\
The gravitational mechanisms, such as galaxy mergers \citep{Toomre1972}, harassment by nearby galaxies and/or the cluster potential \citep{Moore1996}, tidal stripping \citep{Moore1999}, and strangulation \citep{LTC1980} can alter both galaxy morphology and SF activity. The hydrodynamical processes directly act on the gas component of the cluster galaxies through interactions between the hot ($T \sim 10^7 \ K$) intracluster medium (ICM) and the interstellar medium (ISM) --- ram-pressure stripping \citep[RPS;][]{GG1972}, viscous stripping \citep{Nulsen1982}, and thermal evaporation \citep{Cowie1977}. Simulations indicate that RPS can trigger SF within cluster galaxies by compressing the gas \citep{Fujita1999b, Bekki2003, Bekki2010, Roediger2014} and then quench galaxies by removing the gas that would otherwise be available for future SF. RPS may occur rapidly \citep{Domainko2006, Kapferer2009} or gradually over a longer timescale \citep[aka. strangulation; ][]{LTC1980}. Unlike the gravitational interactions, hydrodynamical mechanisms have little effect on the overall stellar distribution \citep{Cortese2021}.
\\
\\
Compared to relaxed clusters (i.e., virialised), merging clusters harbour more extreme environments. During core passage, the relative speed of the merging haloes can reach up to 4500 \kms \citep{Sarazin2002, Markevitch2002, Markevitch2004}, which generates high-speed shocks travelling through the ICM. Because the pressure of the ICM in the region of the shock is higher than the overall ICM pressure, RPS could favour SF more easily \citep{Bekki2003, Bekki2010, Roediger2014, Ruggiero2019}. Therefore, additional impacts on galaxy properties are expected \citep[post-processing;][]{Vijayaraghavan2013}. That said, contradictory results have been presented by authors on whether cluster mergers promote or quench SF. The most important open problems in the field are discussed in the following paragraphs.   
\\
\\
\textit{\textbf{Do cluster mergers trigger star formation?}} Increased global SF activity in merging clusters has been widely reported. Clusters with multiple components, indicative of ongoing dynamical activity, show higher fractions of SFGs relative to the single-component clusters \citep{Cohen2014, Cohen2015, Yoon2020}. Similarly, \cite{Aldas2024}, based on clusters selected in the \textsc{\texttt{Illustris TNG}} simulation, found that disturbed clusters host more blue and SFGs than their relaxed counterparts. \cite{Stroe2017} found an overdensity of \ha emitters in clusters hosting radio shocks that are indicators of ongoing merger activity. On smaller scales, mergers appear to enhance SF as well. \cite{Cortese2004} reported three infalling, highly star-forming subgroups in the young binary cluster A1367. Galaxies between merging subclusters have been observed to exhibit SF and nuclear activity \citep{Ferrari2005, HL2009}. \cite{Owers2012} identified multiple jellyfish galaxies near the merging features in A2744 as an indication of enhanced SF activity due to the merger. \cite{Stroe2015} found unusually high star formation rates (SFR) and fractions of \ha emitters in the region affected by merger-induced shocks in the young post-merger Sausage cluster, where the SFR density is comparable with that seen at cosmic noon ($z\sim$2). In the multiple cluster system A901/2, \cite{RomanOliveira2019} identified 70 jellyfish galaxy candidates with specific star formation rates (sSFR) indicating an enhancement. \cite{Ruggiero2019} conducted a simulation-based analysis on the same system, which showed jellyfish galaxies are preferentially formed near the leading edges of interacting halos due to the boosted ram pressure. Finally, a recent study by \cite{Lourenco2023} reported a mild, yet marginally significant, enhancement in the RPS candidates in interacting clusters in the Wide-Field Nearby Galaxy-Cluster Survey \citep[WINGS;][]{Cava2009} sample.
\\
\\
\textit{\textbf{Do cluster mergers have any effect on star formation?}} Interestingly, merging clusters are also found to be efficient in suppressing SF activity. Numerical simulations reveal an increase in gas mass loss due to RPS during major mergers, leading to weakening/quenching of SF activity \citep{Fujita1999a, Domainko2006, Kapferer2009}. Passive spiral/disk galaxies were found in the centres of merging clusters (i.e., between their brightest cluster galaxies, BCGs) by \cite{Pranger2014} and \cite{Kelkar2020}, suggesting merger-driven quenching. \cite{Deshev2017} found neither recent nor ongoing SF in members in the core of the merger A520, suggesting a scenario of rapid quenching without a starburst episode. In merging \textsc{\texttt{Illustris TNG}} clusters, \cite{Li2023} concluded that merger-induced shocks lead to rapid quenching of SF in low-mass satellites, which is also supported by \cite{Roberts2024} for nine nearby clusters hosting radio relics. There are also studies finding no significant enhancement in SF or quenching in merging systems \citep{Kleiner2014, Wittman2024, Astudillo2025}.
\\
\\
Across previous studies, a bimodal approach emerges: studies either focusing on a single cluster with well-defined dynamics \citep{Owers2012} or on a large cluster sample \citep{Cohen2014}, where dynamical activity is less well-characterised. In this paper, we use a sample of four clusters from the SAMI Galaxy Survey \citep[hereafter SAMI-GS;][]{Bryant2015} where dynamical states are well-characterised. We therefore adopt an approach that is a compromise between the two that have been used so far. This enables a homogeneous analysis of galaxy populations in clusters that have a range of dynamical conditions.
\\
\\
The paper is organised as follows: Section~\ref{sec:Cluster Sample} and Section~\ref{sec:Galaxy Sample} introduce the cluster and galaxy samples, and the spectroscopic data used in this paper. In Section~\ref{sec:Spectral classification}, we outline the spectroscopic classification scheme to define the galaxy populations across the sample. Section~\ref{sec:Results} presents our results, and Section~\ref{sec:Discussion} interprets them in the context of previous studies and simulations. Section~\ref{sec:Conclusion} highlights the key findings in this study. Throughout this paper, we adopt a flat $\Lambda$CDM cosmology with $H_0 = 70 \ km \ s^{-1}\ Mpc^{-1}, \ \Omega_{m} = 0.3, \ \Omega_\Lambda = 0.7$.


\section{Cluster Sample}\label{sec:Cluster Sample}
In this study, we use a sample of clusters from the SAMI-GS \citep{Bryant2015}. As outlined in \citet{Owers2017}, the sample contains eight low-redshift clusters - namely A85, A119, A168, A2399, A3880, A4038, APMCC 0917, and EDCC 0442- spanning a redshift range of $0.029< z < 0.058$, selected from the catalogue of \cite{dePropris2002} in the 2$^\circ$ Field Galaxy Redshift Survey \citep[\texttt{2dFGRS};][]{Colless2001} fields and the Cluster Infall Regions in the Sloan Digital Sky Survey, herefter \texttt{SDSS}, \citep[\texttt{CIRS};][]{Rines2006}. The clusters were chosen to fall within the mass range of $\log(M_{200}/M_\odot) = 14.25 - 15.19$, yet they show differences in their dynamical states. This selection enables us to study how the different merger stages of clusters influence the galaxy properties. Consistent with this scientific motivation, we use a subsample containing four clusters - two merging (A168, A2399) and two relaxed clusters (A3880, EDCC 0442). Section~\ref{subsec:Notes on individual clusters} briefly summarises the literature on the dynamical states of the individual clusters in the sample.

\begin{figure*}[!ht]
    \centering
    \includegraphics[width=\textwidth]{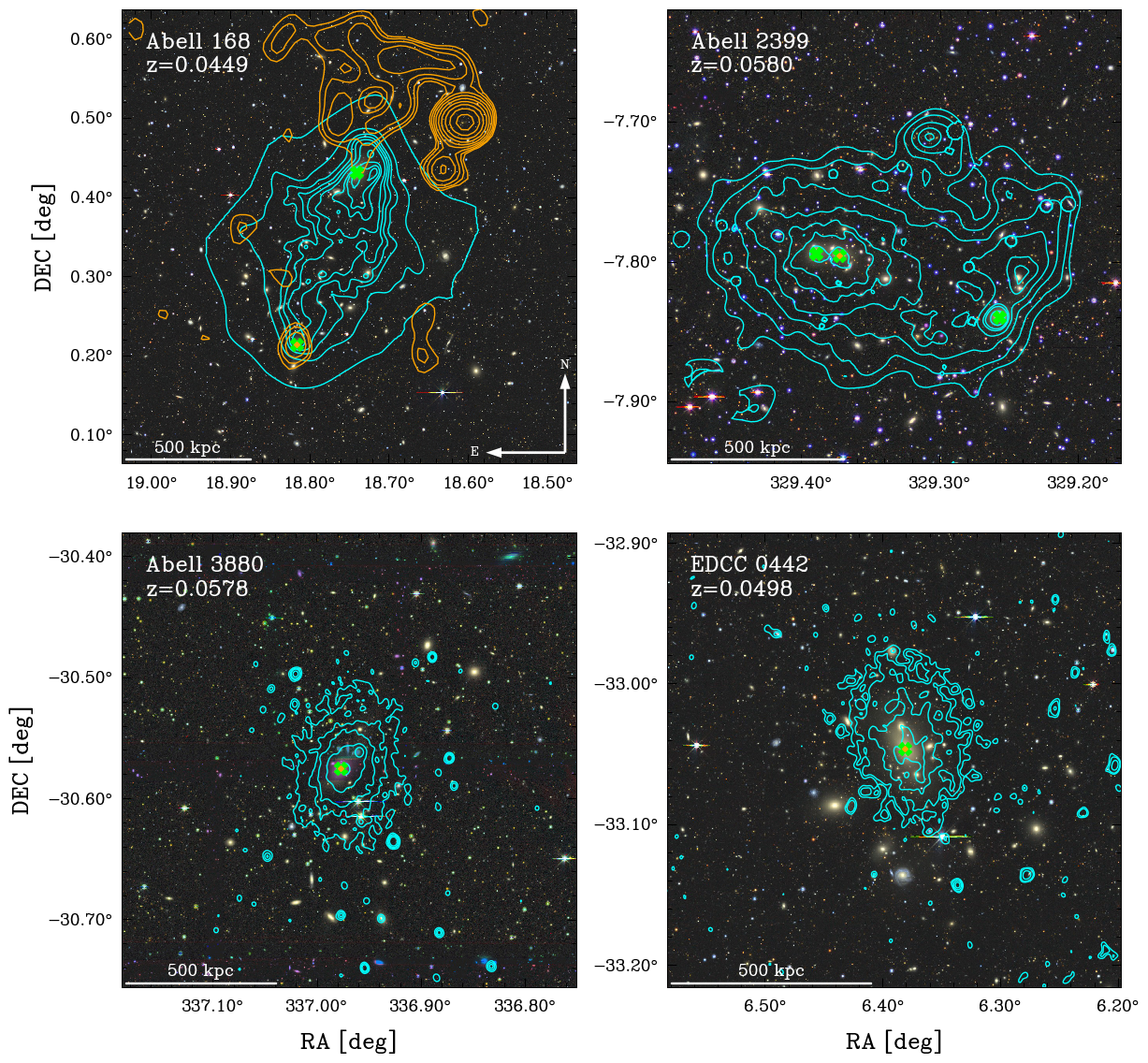}
    \caption{Legacy Survey imaging \citep{Dey2019} of the cluster sample used in this study, overlaid with X-ray emission (cyan contours) from XMM-Newton for A2399 and EDCC 0442 and Chandra for A168 and A3880. The top and bottom panels highlight the merging and relaxed clusters, respectively, with names and redshifts written in the upper left of each panel. In the top left panel, the gold contours show radio emission (170-231 MHz) from the GLEAM survey \citep{Hurley-Walker2017}. The lime crosses and gold diamonds denote the BCGs and the adopted centres of each system, respectively. As shown by the arrows in the top left plot, the orientation is North up, East to the left, and this applies to all clusters. It is evident that X-ray and radio properties of merging clusters differ markedly from those of the relaxed clusters, highlighting the ongoing dynamical activity.}
    \label{Xray_maps}
\end{figure*}

\subsection{Notes on individual clusters}\label{subsec:Notes on individual clusters}
\textbf{A168} (z=0.0449) is a well-studied, off-axis cluster merger occurring nearly on the plane of the sky \citep{Ulmer1992, Yang2004_a, Yang2004_b, Hallman2004, HL2009, Fogarty2014, Owers2017, Dwarakanath2018}. As shown in the top left panel of Figure~\ref{Xray_maps}, it exhibits elongated X-ray emission with two peaks separated by \s700 kpc in projection, aligned along the NW-SE merger axis. The X-ray peaks are centred on the BCG of each subcluster and are offset from the galaxy density peak. \cite{Yang2004_a, Yang2004_b} infer a crossing time-scale of \s0.6 Gyr based on the spatial separation, indicating an early-stage post-merger with a mass ratio, estimated via simulations, between 1:1 and 1:3. \cite{Hallman2004} identified a cold front situated ahead of the northern peak, pointing to a more evolved merger stage. \cite{HL2009} used a linear two-body model to refine the merger timescale. Combining the observational indicators available --- only in X-ray without existing radio observation --- at that time, they favoured a scenario where core-passage occurred \s5 Gyr ago (for our adopted cosmology) and the subclusters are approaching for a second passage. A recent study by \cite{Dwarakanath2018} identified twin radio relics near the northern X-ray peak, indicating a merger-induced shock, with an estimated merger timescale of \s0.5 Gyr.
\\
\\
\textbf{A2399} (z=0.0580) is a dynamically young system exhibiting low surface brightness and strongly bimodal X-ray emission with peaks \s600 kpc apart \citep{Bohringer2007, Bohringer2010, Owers2017} as presented in the top right panel of Figure~\ref{Xray_maps}. The X-ray peaks are coincident with the BCGs and the galaxy density peaks \citep{Fogarty2014}. Moreover, \cite{Fogarty2014} identified extended X-ray emission \citep[referred to as the "plume" in][]{Fogarty2014} extending northward from the western X-ray peak, indicating ongoing merging activity. Substructure analyses of A2399 reveal contradictory results. Using WINGS data, \cite{Flin2006}, \cite{Ramella2007} and \cite{Moretti2017} could not detect any substructures in A2399. In contrast, \cite{Owers2017} detected the presence of substructures through multidimensional (i.e., velocity, spatial, and combined) analysis. One of these substructures is coincident with the western X-ray structure, and was also detected by \cite{Lourenco2020} via 3D (i.e., spatial+velocity) analysis. The X-ray properties of A2399 reveal contact discontinuities in both temperature and brightness profiles, as well as a high central entropy and metallicity \citep{Mitsuishi2018, Lourenco2020}. However, the former authors interpret these features as evidence of a cold front, whereas the latter authors consider the feature as a shock front. Regardless, both suggest that A2399 might have undergone a recent head-on collision (i.e., post-core-passage phase) on the plane of the sky along the NE-SW direction. Yet, neither study considers the plume identified in \cite{Fogarty2014} in their merger scenario. In this paper, we consider an alternative merger scenario where the plume is a result of RPS of gas from the western substructure. The offset of the plume to the NW indicates that the western clump is travelling towards the SE. This scenario suggests that A2399 is a pre-merger system right before the first core passage.
\\
\\
In contrast, A3880 (z=0.0578) and EDCC 0442 (z=0.0498) show no evidence of ongoing merger activity \citep{Owers2017, Lourenco2023}, having relatively regular, symmetric X-ray emission detected with Chandra and XMM-Newton, respectively, well-centred with the BCGs as illustrated in the bottom row of Figure~\ref{Xray_maps}. 

\section{Data and Galaxy Sample}\label{sec:Galaxy Sample}

\begin{figure*}[!t]
    \centering
    \includegraphics[width=\textwidth]{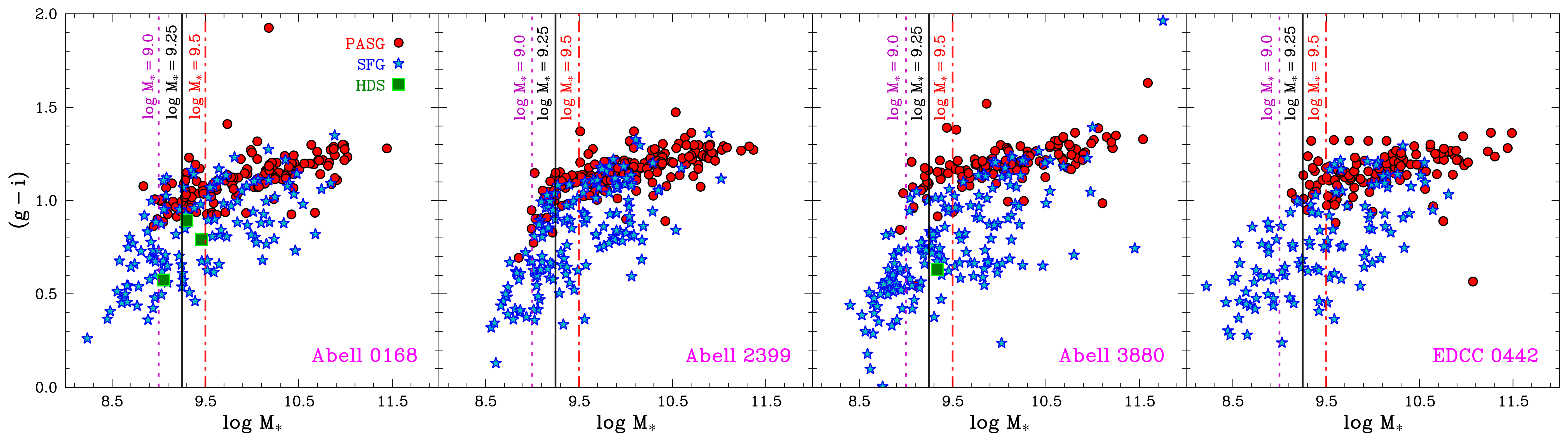}
    \caption{Colour-mass diagram of members for each cluster. The red dots, cyan stars, and green squares represent the galaxies classified as "passive", "star-forming", and "\hd-strong" based on the classification scheme outlined in Section~\ref{sec:Spectral classification}. The magenta dashed, the black solid, and the red dashed lines represent the mass limits for log$(M_\ast/M_\odot)$ = 9.00, 9.25, and 9.50, respectively.}
    \label{CMD_subsample}
\end{figure*}

\subsection{Spectroscopic Data}
In this study, we exploit the single-fibre spectroscopy from the SAMI Cluster Redshift Survey (hereafter SAMI-CRS; \cite{Owers2017}) and SDSS-III \citep{Ahn2014, Alam2015}. The SAMI-CRS was conducted as a part of the SAMI-GS. The survey gathered spectra using the \texttt{2dF/AAOmega} multi-object spectrograph mounted at the prime focus of the 3.9-m Anglo-Australian Telescope over seven nights in 2013. The \texttt{2dF} instrument contains 392 fibres, each with 2 arcsec diameter, spread over a two-degree field of view. These fibres feed light to the \texttt{AAOmega} dual-beam spectrograph \citep{Sharp2006}. The SAMI-CRS targets were observed with the low-resolution 580V and 385R gratings for the blue and red arms, respectively, with a 5700 \AAs dichroic beam splitter. The setup covers a wavelength range of 3700-5850 \AAs in the blue arm and 5600-8800 \AAs in the red arm, with 3.53 \AAs and 5.32 \AAs full width at half-maximum (FWHM) resolution for the blue and red arms, respectively.
\\
\\
In total, 22066 spectra were taken during the SAMI-CRS. The details of the redshift measurements, the quality assessment (i.e., accuracy and precision), and duplicate analysis are given in \citet{Owers2017}. Based on the quality measurements given in \citet{Owers2017}, we only consider the SAMI-CRS spectra with reliable redshift measurements (i.e., $\textrm{confidence}>0.9$), yielding 7648 spectra including duplicates for our subsample (out to \s4.5 \rtwo). SDSS provides 492 spectra for A168 and A2399 out to similar cluster-centric radii.

\subsection{Galaxy Sample}\label{Galaxy Sample}
The SAMI-CRS provides a high completeness in redshifts measurements (\s94\%) within 2 \rtwo for Petrosian $r$-band magnitudes $r_{\rm petro}\leq19.4$ \citep{Owers2017}, by combining \texttt{AAOmega} observations and a compilation of pre-existing redshifts from other surveys, such as \texttt{SDSS} \citep{Ahn2014}, \texttt{2dFGRS} \citep{Colless2001}, \texttt{6dFGS} \citep{Jones2009}, \texttt{WINGS} \citep{Cava2009}, \texttt{CAIRNS} \citep{Rines2003}, \texttt{NFPS} \citep{Smith2004}, \texttt{ENACS} \citep{Katgert1996}, and a catalogue for A85 from \cite{Durret1998}. Cluster membership is determined via an iterative procedure based on these redshifts and the spatial positions of galaxies as described in \cite{Owers2017}. The global cluster properties (i.e.,$\sigma_{200}$, \rtwo) and the dynamical characterisation (e.g., presence of any substructures) are determined considering these confirmed members. This procedure \citep[detailed in][]{Owers2017} results in a membership catalogue containing 1167 cluster galaxies for A168, A2399, A3880, and EDCC 0442 out to 2 \rtwo.
\\
\\
Based on the member galaxies, we select our galaxy sample for the analysis outlined in Section~\ref{sec:Spectral classification} by applying two main criteria. As a first step, we only include members within 2 \rtwo having either SAMI-CRS spectra with reliable redshifts (i.e., $\textrm{confidence}>0.9$) or SDSS spectra, removing 27 galaxies. Another key consideration is probing the same stellar mass range across the sample, which required homogeneous photometric measurements for stellar mass estimation. The SAMI-CRS photometry comes from \texttt{SDSS-DR9} \citep{Ahn2012} and \texttt{VST/ATLAS} \citep{Shanks2015}, and the homogenisation procedure is outlined in Section 5 of \cite{Owers2017}. There are 1103 members with available photometry. For these objects, the stellar masses are estimated using the empirical relation between (g-i) colour and (M/L)$_i$ given by \cite{Taylor2011}, also outlined in \citet{Bryant2015} and \citet{Owers2017} for apparent magnitudes. The colour-magnitude diagram shown in Figure 7 of \cite{Owers2017} and colour-mass diagram given in Figure~\ref{CMD_subsample} indicate that the low-mass regime is incomplete in terms of red sequence galaxies, especially for EDCC 0442. This could introduce a bias toward blue cloud galaxies. Therefore, we apply a cut on the stellar masses as $\log(M_\ast/M_\odot)\geq9.25$. This yields 829 galaxies --- $N_{\text{A168}}=193$, $N_{\text{A2399}}=241$, $N_{\text{A3880}}=211$, and $N_{\text{EDCC 0442}}=184$.

\section{Spectroscopic Classification} \label{sec:Spectral classification}
Cluster mergers that trigger SF in their resident galaxies are expected to produce a higher incidence of SFGs \citep{Stroe2015} with respect to their relaxed counterparts. In the case of a boost in rapid quenching of SF activity, a notable fraction of recently quenched galaxies with strong \hd signatures is expected \citep{Couch1987, Poggianti1999, Owers2019} to emerge in these environments. To address this, one needs to identify these populations by examining their spectral features. In this section, we outline the line measurements and the definition of populations through a spectroscopic classification scheme. 

\subsection{Emission Line Measurements}\label{Emission line measurements}
We measure emission line properties by employing the penalized PiXel Fitting code \citep[\ppxf;][]{Cappellari2004,Cappellari2017}. \ppxf allows us to simultaneously fit the emission lines along with the stellar kinematics and stellar populations. In order to measure emission line properties accurately, the underlying stellar continuum must be modelled and subtracted. As recommended in Section 3.6 of \cite{Cappellari2017} and implemented by \citet{Sande2017}, \citet{Owers2019}, and \citet{Oh2020}, this can be achieved by iteratively running \ppxf. The steps adopted here are described below:

\begin{itemize}[leftmargin=0.1in]
    \item Initially, the spectrum and noise are fed into \ppxf, excluding bad pixels. Instead of masking the emission lines, we also simultaneously fit gas templates to account for the contribution of the emission features to the continuum. We only use an additive polynomial with a degree of 12 (\texttt{degree=12, mdegree=-1}) to minimise mismatches between theoretical \texttt{MILES} templates \citep{Sanchez2006, Vazdekis2010, Falcon2011} and the spectrum. The original noise is then rescaled based on the residual between the spectrum and the best fit and used in the next step. 
    
    \item The second run involves the removal of $3\sigma$ outliers, using the CLEAN parameter. The best-fit solution for the stellar kinematics is kept as an initial guess for the third iteration. 
    
    \item In the final run, we use a 12th-degree multiplicative polynomial in place of the additive polynomial (\texttt{degree=-1, mdegree=12}). 
\end{itemize}
Throughout the procedure, we assume the same kinematics for all the emission lines. 
\\
\\
We measure the equivalent widths (EW) of \ha and \nii using the results from the \ppxf procedure outlined above. Firstly, a continuum band is defined on either side of the \ha and \nii lines to determine the continuum level from the best-fit stellar continuum, and then \ha and \nii fluxes are divided by that continuum level to measure their EWs. A similar approach is also adopted for the absorption line measurements discussed in the next section.

\subsection{Absorption Line Measurements}\label{Absorption line measurements}
The \hd absorption line equivalent width, EW(\hd), is measured from the best-fit stellar continuum from pPXF in Section~\ref{Emission line measurements} using the procedure described in \cite{Cardiel1998}. We estimate EW(\hd) and the associated uncertainties using a Monte Carlo (MC) routine, performing 1000 iterations (i.e., $\rm N_{iteration} = 1000$) of resampling on the best-fit stellar continuum. For each iteration, we perturb the best-fit continuum by adding a random value sampled from a normal distribution using the noise spectrum as the standard deviation at each spectral pixel. Subsequently, absorption line properties are measured in the observed frame; equivalent width and its uncertainty are defined as the mean and standard deviation of the distribution of EW values measured in each MC iteration.
\subsection{Classification Scheme}\label{Classification}
Once both emission and absorption line properties are measured, we are able to conduct a spectral classification to define galaxy populations. For this purpose, we adopt a similar classification scheme to that outlined in \cite{Owers2019}. 
\\
\\
\textbf{\textit{Emission - }}We define an emission spectrum as one in which at least two emission lines are detected with SNR \gt 3: one must be a primary line (e.g., \ha or \nii) and the second one must be one of \ha, \nii, \hb, \oiii. Here, SNR is the ratio between the flux and its uncertainty (i.e., $flux/\sigma_{flux}$). We employ two additional criteria that help to remove spurious low-amplitude emission line detections that occur due to template mismatch during the pPXF fitting process. First, we only include those spectra where EW(\ha or \nii) \gt 1 \AA. Second, we only include spectra with reliably measured gas kinematics, i.e., where $\sigma_{gas} < 500$ \kms and $|v_{gas}|<500$ \kms.
\\
\\
The ionising source of emission line galaxies is classified as follows:
\begin{itemize}[leftmargin=0.1in]
    \item For emission-line galaxies where all of the lines \ha, \nii, \hb, \oiii are detected, we use the BPT diagram \citep{BPT1981} and the populations are defined based on the demarcation line descriptions from \cite{Kewley2001} (Ke01) and \cite{Kauffmann2003} (Ka03); the line ratios of SFGs (SF) lie below and left of the dividing line of Ka03, intermediate (INT) galaxies lie between that and the dividing line of Ke01, while non-SFGs (NSF) lie above and right of it.
    
    \item For galaxies without \hb and/or \oiii, we use the WHAN diagram of \cite{CidFernandes2010}, in which emission line classification is done using the EW(\ha) and the ratio between \nii and \ha. The demarcation lines to separate SF, INT, and NSF galaxies are log(\nii/\ha) = -0.32 and log(\nii/\ha) = -0.1, which are approximations to the lines of Ka03 and Ke01, respectively.

    \item Galaxies not presenting both primary lines are classified as SF if \ha is present. Otherwise, when only \nii is detected, they are classified as NSF.

    \item Depending on the value of EW(\ha), sub-classification follows \cite{CidFernandes2010}. If EW(\ha) \lt 3 \AA, the galaxy can be classified as \textit{w}SF or \textit{r}INT, or \textit{r}NSF. NSF spectra with 3 \AAs\lt EW(\ha) \lt 6 \AA\ are \textit{w}NSF, while those with EW(\ha) \gt 6 \AAs are \textit{s}NSF. The prefixes \textit{"r/w/s"} denote \textit{"retired/weak/strong"}.

\end{itemize}
\textbf{\textit{Absorption - }} We consider the spectra that do not satisfy the criteria given above for absorption line classification. We classify them based on their absorption line properties, following the criteria from \citet{Owers2019} as follows:
    \begin{itemize}[leftmargin=0.1in]
        \item The spectra with $SNR(H\delta) > 3$, where $SNR = |EW(H\delta)|/\sigma_{EW}$, and $EW(H\delta)<-3$ \AA\xspace are classified as \hd-strong (i.e., HDS).
        \item If $EW(H\delta)>-3$ \AA, they are classified as having passive spectrum.
    \end{itemize}

\noindent \textbf{\textit{Passive versus star-forming classes - }}Combining all spectral information from both emission and absorption line diagnostics, we divide cluster galaxies into passive galaxies (PASG), SFGs, and \hd-strong galaxies (HDSG). PASGs include galaxies with spectra classified as passive, \textit{r/w/s}NSF, \textit{r}INT, \textit{w}SF; SFGs are the remaining combination of SF and INT galaxies; HDSGs are those labelled as HDS. 

\section{Results}\label{sec:Results}
\subsection{Population fractions}\label{Result - Fractions}
\begin{figure}[!b]
    \centering
    \includegraphics[width=\linewidth]{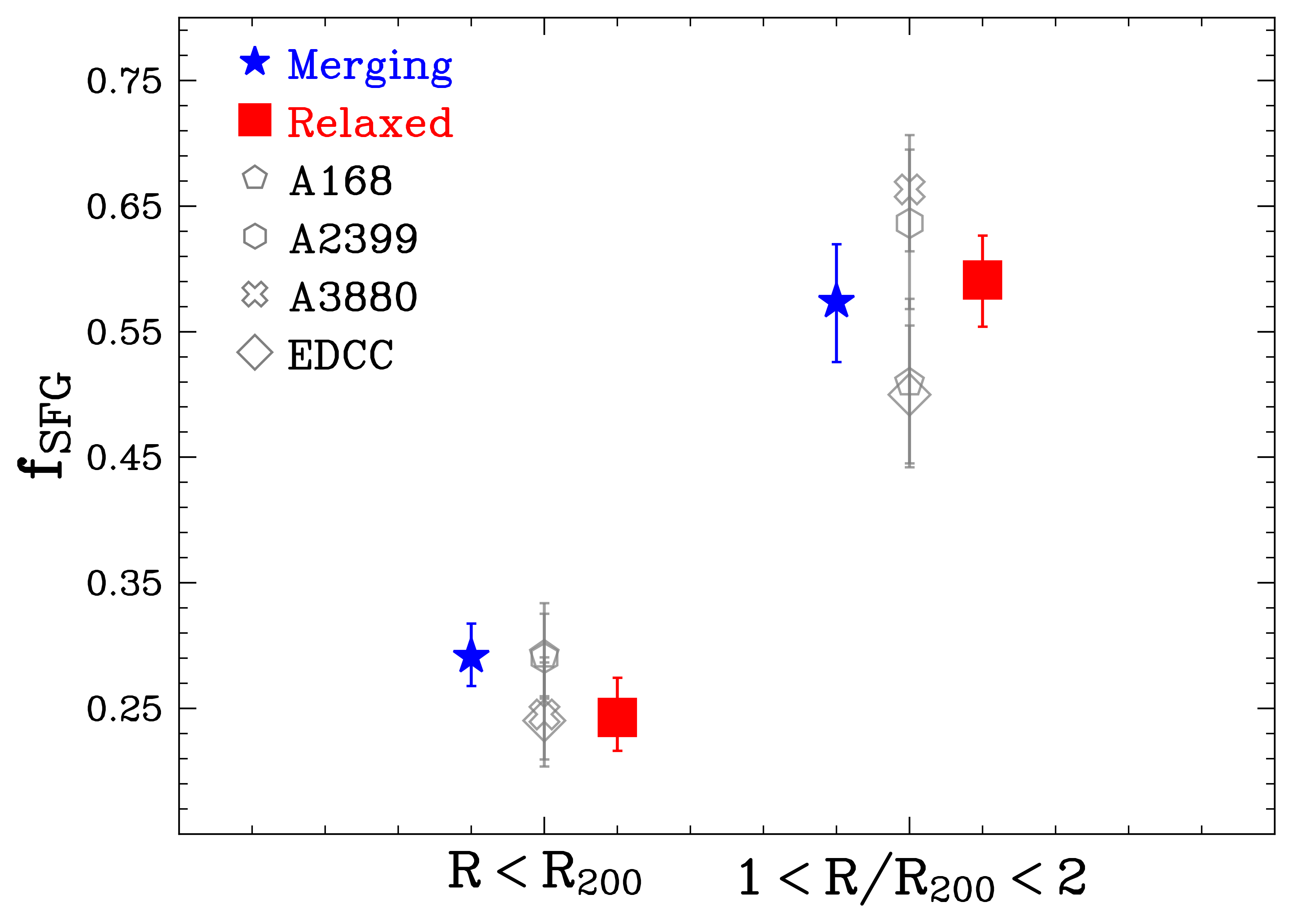}
    \caption{Star-forming galaxy fractions for different cluster-centric regions - R \lt\rtwo and 1 \lt R/\rtwo \lt 2. The grey open symbols (i.e., pentagon, hexagon, thick cross, and diamond) show each cluster's fraction (i.e., A168, A2399, A3880, and EDCC 0442, respectively). The blue stars and the red squares represent the stacked samples based on similar cluster dynamics (i.e., merging and relaxed, respectively). The error bars show the uncertainties on the fractions estimated through the approach given in \citet{Cameron2011}.}
    \label{SFG_fractions}
\end{figure}

\rotatebox[origin=c]{90}{}
\begin{table*}[!ht]
    \centering
    \renewcommand{\arraystretch}{1.5}
    \begin{NiceTabular}{|c|c|c|c|c|c|c|c|c|}[corners=NW]
            \Hline
            & & \Block{1-7}{Counts \& Fractions} \\
            \Hline
            Region & Clusters & $\mathrm{N_{Total}}$ &  $\mathrm{N_{PASG}}$ & $\mathrm{N_{SFG}}$ & $\mathrm{N_{HDSG}}$ & $\mathrm{f_{PASG}}$ &  $\mathrm{f_{SFG}}$ & $\mathrm{f_{HDSG}}$ \\
            \Hline
            \Block{6-1}{$\mathrm{R<R_{200}}$} & Abell 168 & 140 & 97 & 41 & 2 & $\mathrm{0.693_{0.041}^{0.036}}$ & $\mathrm{0.293_{0.035}^{0.041}}$ & $\mathrm{0.014_{0.005}^{0.018}}$ \\
            & Abell 2399 & 186 & 132 & 54 & 0 & $\mathrm{0.710_{0.035}^{0.031}}$ & $\mathrm{0.290_{0.031}^{0.035}}$ & $\mathrm{0.000_{0.001}^{0.010}}$ \\
            & Abell 3880 & 110 & 82 & 27 & 1 & $\mathrm{0.745_{0.045}^{0.037}}$ & $\mathrm{0.245_{0.036}^{0.045}}$ & $\mathrm{0.009_{0.003}^{0.020}}$ \\
            & EDCC 0442 & 104 & 79 & 25 & 0 & $\mathrm{0.760_{0.046}^{0.037}}$ & $\mathrm{0.240_{0.037}^{0.046}}$ & $\mathrm{0.000_{0.002}^{0.017}}$ \\
            \Hline
            & \textbf{Merging} & 326 & 229 & 95 & 2 & $\mathrm{0.702_{0.026}^{0.024}}$ & $\mathrm{0.291_{0.024}^{0.026}}$ & $\mathrm{0.006_{0.002}^{0.008}}$\\
            & \textbf{Relaxed} & 214 & 161 & 52 & 1 & $\mathrm{0.752_{0.032}^{0.027}}$ & $\mathrm{0.243_{0.027}^{0.031}}$ & $\mathrm{0.005_{0.001}^{0.011}}$\\
            \Hline   
            \Block{6-1}{$\mathrm{1 < R/R_{200} < 2}$} & Abell 168 & 53 & 26 & 27 & 0 & $\mathrm{0.491_{0.067}^{0.067}}$ & $\mathrm{0.509_{0.067}^{0.067}}$ & $\mathrm{0.000_{0.003}^{0.033}}$ \\
            & Abell 2399 & 55 & 20 & 35 & 0 & $\mathrm{0.364_{0.059}^{0.068}}$ & $\mathrm{0.636_{0.068}^{0.059}}$ & $\mathrm{0.000_{0.003}^{0.032}}$ \\
            & Abell 3880 & 101 & 34 & 67 & 0 & $\mathrm{0.337_{0.043}^{0.050}}$ & $\mathrm{0.663_{0.050}^{0.043}}$ & $\mathrm{0.000_{0.002}^{0.018}}$ \\
            & EDCC 0442 & 80 & 40 & 40 & 0 & $\mathrm{0.500_{0.055}^{0.055}}$ & $\mathrm{0.500_{0.055}^{0.055}}$ & $\mathrm{0.000_{0.002}^{0.022}}$ \\
            \Hline 
            & \textbf{Merging} & 108 & 46 & 62 & 0 & $\mathrm{0.426_{0.046}^{0.048}}$ & $\mathrm{0.574_{0.048}^{0.046}}$ & $\mathrm{0.000_{0.002}^{0.017}}$\\
            & \textbf{Relaxed} & 181 & 74 & 107 & 0 & $\mathrm{0.409_{0.035}^{0.037}}$ & $\mathrm{0.591_{0.037}^{0.035}}$ & $\mathrm{0.000_{0.001}^{0.010}}$\\
            \Hline
    \end{NiceTabular}
    \caption{Population table for galaxies for each cluster and stacked cluster sample within different cluster-centric portions. Errors are estimated based on the recipe given by \citet{Cameron2011}.}
    \label{fraction_table}
\end{table*}

\noindent As a first step towards understanding if the merger activity in A168 and A2399 is affecting the star-forming properties, we compare the fractions of different spectroscopic classes to those in the relaxed sample. If mergers are responsible for triggering SF, we expect a higher fraction of SFGs in merger-affected regions \citep{Ferrari2005, Stroe2015, HL2009}. Alternatively, if mergers are more effective at rapidly quenching galaxies, we expect a higher fraction of HDS galaxies to be found in merger regions. To that end, we define the fraction of SFGs, $f_{SFG}$, in a region as
\begin{equation}
    f_{SFG} = \frac{N_{SFG}}{N_{PASG} + N_{SFG} + N_{HDSG}},
\end{equation}
\noindent where $N_{PASG}$, $N_{SFG}$, and $N_{HDSG}$ denote the number of passive, star-forming, and \hd-strong galaxies, respectively. We measure the fractions within two cluster-centric regions: (i) $\rm R \leq R_{200}$, where the impact of the mergers is expected to be greater, and (ii) the cluster outskirts ($\rm 1 < R/R_{200} \leq 2$), where they are expected to have had less effect. The outskirts may also offer insight into SF activity around merger regions, making them a great benchmark for comparing non-merger-affected regions across the sample.
\\
\\
Figure~\ref{SFG_fractions} and Table~\ref{fraction_table} show the fraction of SFGs within \rtwo and in the range 1 \lt R/\rtwo \lt 2 for our cluster sample. The results for individual clusters are shown as open grey symbols. The blue stars and red squares show the combined results for the merging and relaxed clusters, respectively. Uncertainties represent $1\sigma$ confidence intervals derived from the recipe of \citet{Cameron2011}.
\\
\\
When R \lt \rtwo is considered, the PASGs are the main population regardless of the cluster merger state. PASGs account for \fraction{70.2}{2.4}{2.6} (229/326) of the merging sample, and \fraction{75.2}{2.7}{3.2} (161/214) of the relaxed clusters. SFGs contribute \fraction{29.1}{2.6}{2.4} (95/326) and \fraction{24.3}{3.1}{2.7} (52/214) of the central galaxy population in merging and relaxed clusters, respectively. The $f_{SFG}$ within \rtwo is higher in merging clusters (29.1\%) compared to relaxed clusters (24.3\%), although the difference, as estimated through a two-proportion Z-test, is not significant ($\sim1.2\sigma$). Across the whole sample, only 3 galaxies (i.e., \lsim 1\%) present strong \hd signatures, making up \fraction{0.6}{0.8}{0.2} (2/326) and \fraction{0.5}{1.1}{0.1} (1/214) of merging (only in A168) and relaxed clusters (only in A3880) for R \lt \rtwo, respectively. Due to the low count, we hereafter combine HDSGs and PASGs.
\\
\\
In contrast, the trend seen within \rtwo is reversed at the outskirts (i.e., 1 \lt R/\rtwo \lt 2) where the presence of SFGs becomes more pronounced. Merging and relaxed clusters show similar fractions of PASGs and SFGs without any HDSGs detected. \fraction{42.6}{4.8}{4.6} (46/108) and \fraction{40.9}{3.7}{3.5} (74/181) of galaxies are classified as PASGs, whereas SFGs account for \fraction{57.4}{4.6}{4.8} (62/108) and \fraction{59.1}{3.5}{3.7} (107/181) of outskirt galaxies for each sample, respectively. 

\subsection{The Spatial Distributions of Passive and Star-forming Galaxies}\label{Result - Spatial distribution}
\begin{figure*}[!ht]
    \centering
    \includegraphics[width=\textwidth]{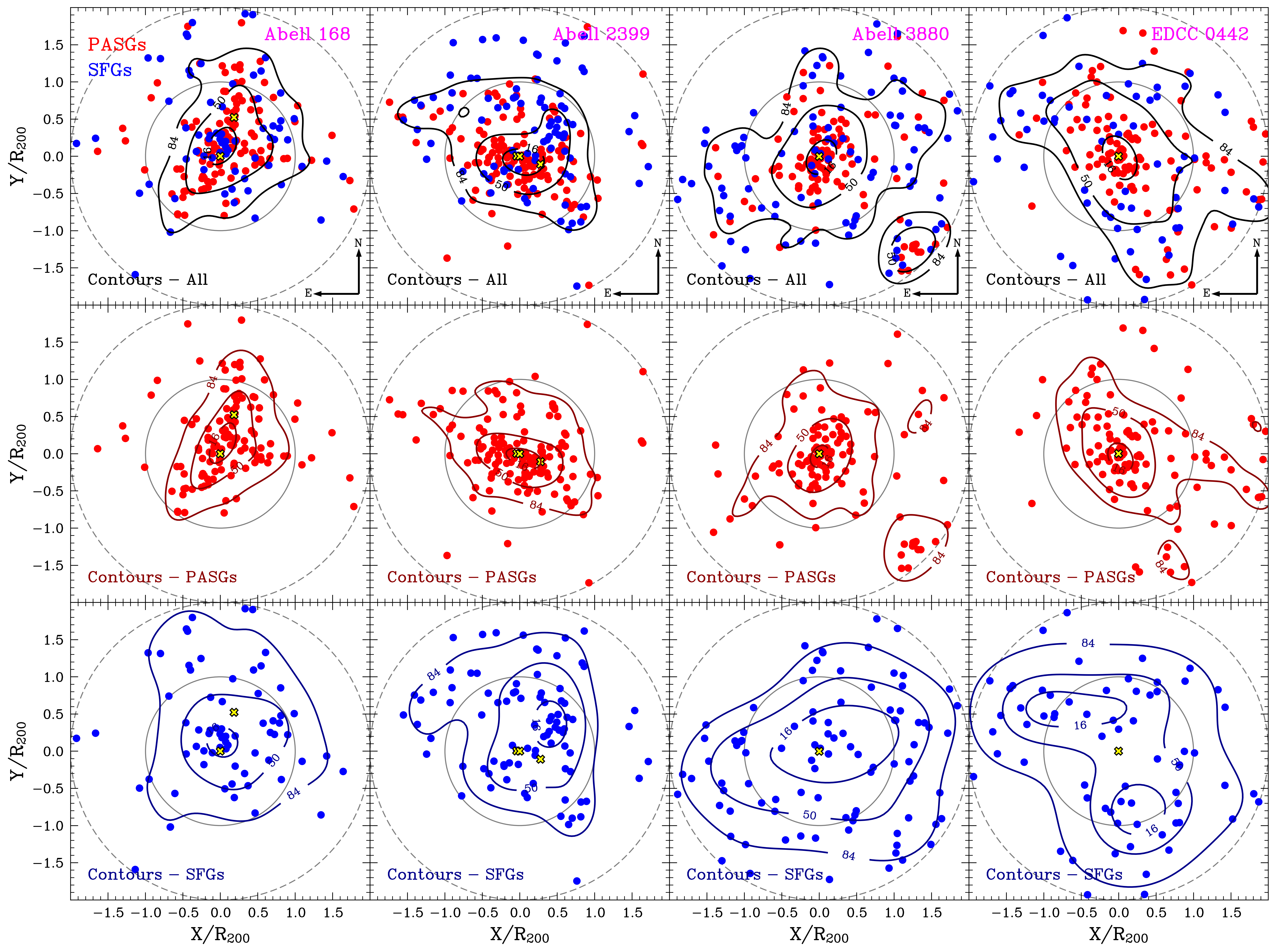}
    \caption{Spatial distribution of cluster galaxies. Columns represent individual clusters, while rows highlight all, passive (red) and star-forming galaxies (blue) from top to bottom, respectively. The $16^{th}, \ 50^{th}, and \ 84^{th}$ percentile contours are generated from the smoothed kernel density estimate (KDE) for the population highlighted in the lower left of each panel using \texttt{ks} package \citep{Duong2012} in \texttt{R}. As indicated by the black arrows in the top row, North is up and East is to the left.}
    \label{Spatial_distribution}
\end{figure*}

In Section~\ref{Result - Fractions}, we found hints that there is a higher $f_{SFG}$ found within \rtwo for the merging clusters when compared to the relaxed sample. To determine if the higher $f_{SFG}$ is related to the merger activity, in this section, we investigate the spatial distributions of galaxies across the sample by highlighting each spectral class.
\\
\\
Figure~\ref{Spatial_distribution} illustrates the spatial distribution of members for each cluster. The red and blue points represent PASGs and SFGs, respectively. The positions of the BCGs are indicated by yellow crosses. From left to right, the columns show the distributions for A168, A2399, A3880, and EDCC 0442, respectively, while from top to bottom each row shows the distribution for one of the galaxy populations, being "all galaxies" (i.e., PASGs+SFGs), "PASGs", and "SFGs". Contours represent the $16^{th}, \ 50^{th}$, and $84^{th}$ percentiles of the distribution for the corresponding population. The contours are generated from the smoothed kernel density estimate (KDE) using the \texttt{ks} package \citep{Duong2007} in \texttt{R}. Below, we describe the results for each cluster.

\textbf{A168:} The overall galaxy distribution is elongated along the merger axis, running NW to SE. This elongation is more pronounced for the PASGs that dominate the region within \rtwo. The elongation is likely due to unresolved substructures, whose presence is revealed by the X-ray data. In contrast, the SFGs do not share the NW-SE elongation that is seen in the PASGs. The distribution of SFGs is more centrally concentrated; the majority of the SFGs (\s60\% - 41/68) are located within \rtwo. There is a dense clump of SFGs just north of the southern BCG (yellow cross) in the region between substructures, consistent with findings of \cite{HL2009}. Two galaxies classified as HDSGs are also found to be in the region between substructures.

\textbf{A2399:} The density peak coincides with two central BCGs. The $50^{th}$ percentile KDE contour extends to the NW, indicating a local excess in the projected galaxy density at $X=0.5, Y=0.5$. The PASGs show a similar spatial distribution to the overall sample, although the elongation is more pronounced, and there is no evidence for a local excess at $X=0.5, Y=0.5$. The elongation of the PASG distribution indicates that there may be an unresolved substructure associated with the third BCG at $X=0.3, Y=-0.1$. As shown in Figure~\ref{Xray_maps}, this third BCG is associated with the X-ray structure. The SFGs present an asymmetric spatial distribution, with the main surface density peak located at (X, Y) = \s(0.5, 0.5) \rtwo.

\textbf{A3880:} As expected for a dynamically relaxed system, A3880 hosts a quite symmetric galaxy distribution. The main galaxy density peak is well-centred around the BCG. There is a secondary peak situated at the outskirts (R\s 2\rtwo) to the SW, at X=1.25, Y=-1.25, likely associated with an infalling group. PASGs mainly populate the central region. The distribution of the SFGs is more extended when compared with those seen in A168 and A2399; the majority (\s70\%) of SFGs are found outside \rtwo. The only HDS galaxy in A3880 is found close to the centre. 

\textbf{EDCC 0442:} The surface density distribution is elongated along a NE-SW axis, and the peak is coincident with the BCG. The distribution of passive galaxies is similar to the overall galaxy distribution, although the SW extension is more prominent in PASGs, indicating accretion from the surroundings. In contrast to A168, A2399, and A3880, SFGs are completely absent from the central \s0.5\rtwo of EDCC 0442. Like A3880, the SFGs are predominantly located on the cluster outskirts.
\\
\\
We can conclude that merging clusters have a more mixed spatial distribution compared to the relaxed systems. PASGs reside in the cluster core regardless of the dynamical state. On the other hand, SFGs in the relaxed clusters have a relatively symmetric spatial configuration, predominantly occupying the outskirts, whereas merging systems host an SFG population that also populates the central \rtwo and exhibits an asymmetric distribution on the sky.


\subsection{Radial Distribution}\label{Result - Radial distribution}
In Section~\ref{Result - Spatial distribution}, we show qualitatively that there appears to be a difference in the radial distribution of SFGs for the merging and relaxed clusters. In order to quantify this result and determine the statistical significance, we look at the distribution of cluster-centric distances per population across the sample.
\begin{figure*}[!t]
    \centering
    \includegraphics[width=.75\textwidth]{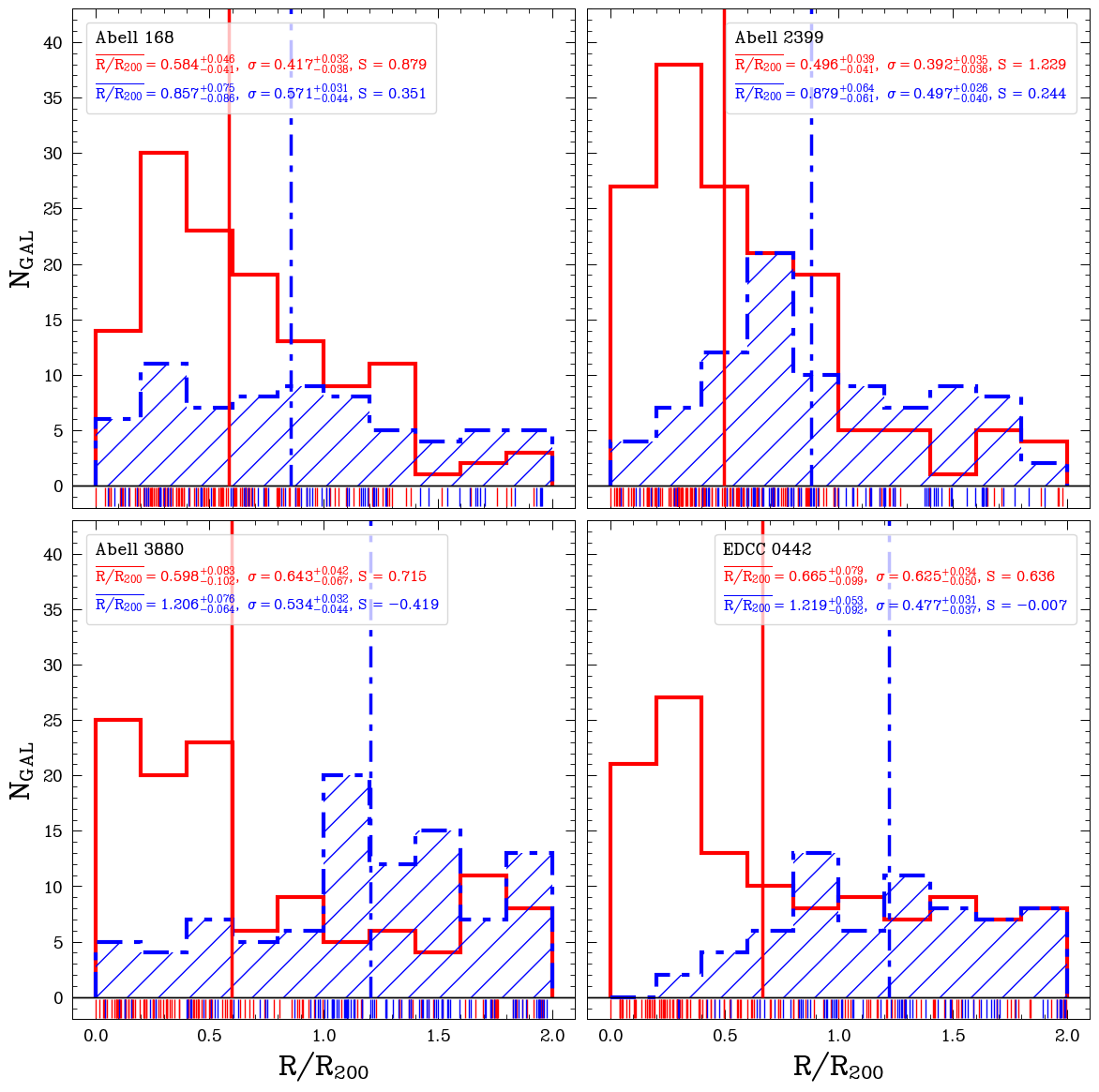}
    \caption{Normalised cluster-centric distance distributions. The red and blue histograms highlight the distributions for PASG and SFGs, respectively. The vertical lines show the mean values for PASGs and SFGs, respectively, calculated via biweight estimation and given in the upper left of each panel alongside the biweighted dispersion and the skewness. The stripe density plots highlight exact values for each galaxy.}
    \label{Cluster-centric distribution}
\end{figure*}
\vspace{2 mm}
\\
Figure~\ref{Cluster-centric distribution} shows the distribution of galaxies as a function of normalised cluster-centric distances (R/\rtwo) across the sample. In each panel of Figure~\ref{Cluster-centric distribution}, the red and blue histograms and vertical lines show the distributions and their mean values calculated via biweight estimation \citep{Beers1990} for PASGs and SFGs, respectively. The uncertainties are estimated through bootstrapping. Regardless of the dynamical states, PASGs are localised in the central \rtwo with a tail toward larger cluster-centric distances. The mean R/\rtwo of the PASGs, shown as a solid red vertical line in each panel of Figure~\ref{Cluster-centric distribution}, ranges from \s 0.5 to 0.7 \rtwo (A168 = 0.584, A2399 = 0.496, A3880 = 0.598, and EDCC 0442 = 0.665). The distributions show positive skewness values (A168 = 0.879, A2399 = 1.229, A3880 = 0.715, and EDCC 0442= 0.636). The number of PASGs per bin for R \gsim 0.5 \rtwo is relatively constant in the relaxed clusters A3880 and EDCC 0442, whereas the merging clusters A168 and A2399 present decreasing numbers of PASGs in the same cluster-centric portion.
\\
\\
In order to quantify how significantly different the PASG distributions are across the sample, we apply the two-sample Kolmogorov-Smirnov \citep[KS;][]{KS} and Anderson-Darling \citep[AD;][]{ADtest} tests on a cluster-to-cluster basis. Both tests return higher p-values (i.e., $p\gg0.05$) for clusters with similar dynamical states (i.e., A168 vs A2399 and A3880 vs EDCC 0442). This means that we cannot reject the null hypothesis that both distributions are drawn from the same parent sample. In contrast, when we compare merging and relaxed clusters, the KS test suggests the opposite situation - p-values are \s0.04, \s0.09 (AD test returns 0.035) for A168 vs A3880 and A168 vs EDCC 0442, respectively, and $p<0.05$ for both A2399 vs A3880 and A2399 vs EDCC 0442, respectively, indicating that two distributions are likely to be drawn from different parent samples.
\\
\\
In the case of SFGs, the R/\rtwo distributions notably differ between merging and relaxed clusters. Figure~\ref{Cluster-centric distribution} shows that the number of SFGs in relaxed clusters increases with increasing cluster-centric distances. On the other hand, the number of SFGs in merging clusters is approximately constant or decreasing towards the outskirts. In particular, A2399 shows a peak in the number of SFGs around \s 0.6 \rtwo. In contrast with that of the PASGs, the mean values of R/\rtwo distributions of the SFGs appear similar for clusters with similar dynamics; \s0.86, and \s0.88 for A168 and A2399, and \s1.21 and \s1.22 for A3880 and EDCC 0442, respectively. Moreover, the skewness values also highlight this difference between dynamical states. A168 and A2399 have positive skewness values as \s0.35 and \s0.24, respectively, whereas relaxed clusters present negative skewness values - \s-0.42 and \s-0.01 for A3880 and EDCC 0442, respectively. 

To determine the significance of the differences in the R/\rtwo distributions of SFGs, we again use the KS-test. Comparing the merging clusters A168 and A2399, the KS-test returns a p-value of \s0.28. Similarly, the KS-test returns a p-value \s0.61 for the comparison of the relaxed clusters A3880 and EDCC 0442. Thus, there is no statistically significant difference between the R/\rtwo distributions for the two clusters within the relaxed sample, nor the two clusters within the merging sample. In contrast, comparing clusters across the merging and relaxed samples yields significant differences; each of the comparisons A168 vs A3880, A168 vs EDCC 0442, A2399 vs A3880 and A2399 vs EDCC 0442 returns a p-value $\ll0.05$. These results indicate that the distributions of R/R200 for SFGs in the merging clusters are significantly different from those in the relaxed clusters in our sample. Moreover, we test the consistency between the means of the R/\rtwo distributions of SFGs using both biweighted metrics and the original means with standard deviations, applying a two-sample t-test. For each merging--relaxed cluster pair, the difference between means exceeds $\sim3.5\sigma$, supporting the results from the KS test.
\\
\\
Lastly, if we stack the clusters based on their dynamical states, the difference in the R/\rtwo distributions for each galaxy population becomes more significant. The mean R/\rtwo values of PASGs are 0.52 and 0.52 for merging and relaxed clusters, respectively, yet the KS test returns a p-value of \s0.002. Regarding SFGs, the mean values are 0.83 and 1.21 for merging and relaxed clusters, and the p-value from the KS test is \s$10^{-7}$. 

\subsection{Projected phase-space Distribution}\label{Result - PPS distribution}
The position of a galaxy in projected phase-space (PPS) is determined using the projected cluster-centric distance and the line-of-sight velocity relative to the cluster redshift.  The position in PPS is a useful diagnostic of cluster membership and also for segregating populations based on their accretion times \citep{Haines2012, Noble2013, Oman2013, Haines2015, Owers2017, Rhee2017, Owers2019}.
\begin{figure*}[t]
    \centering
    \includegraphics[width=\textwidth]{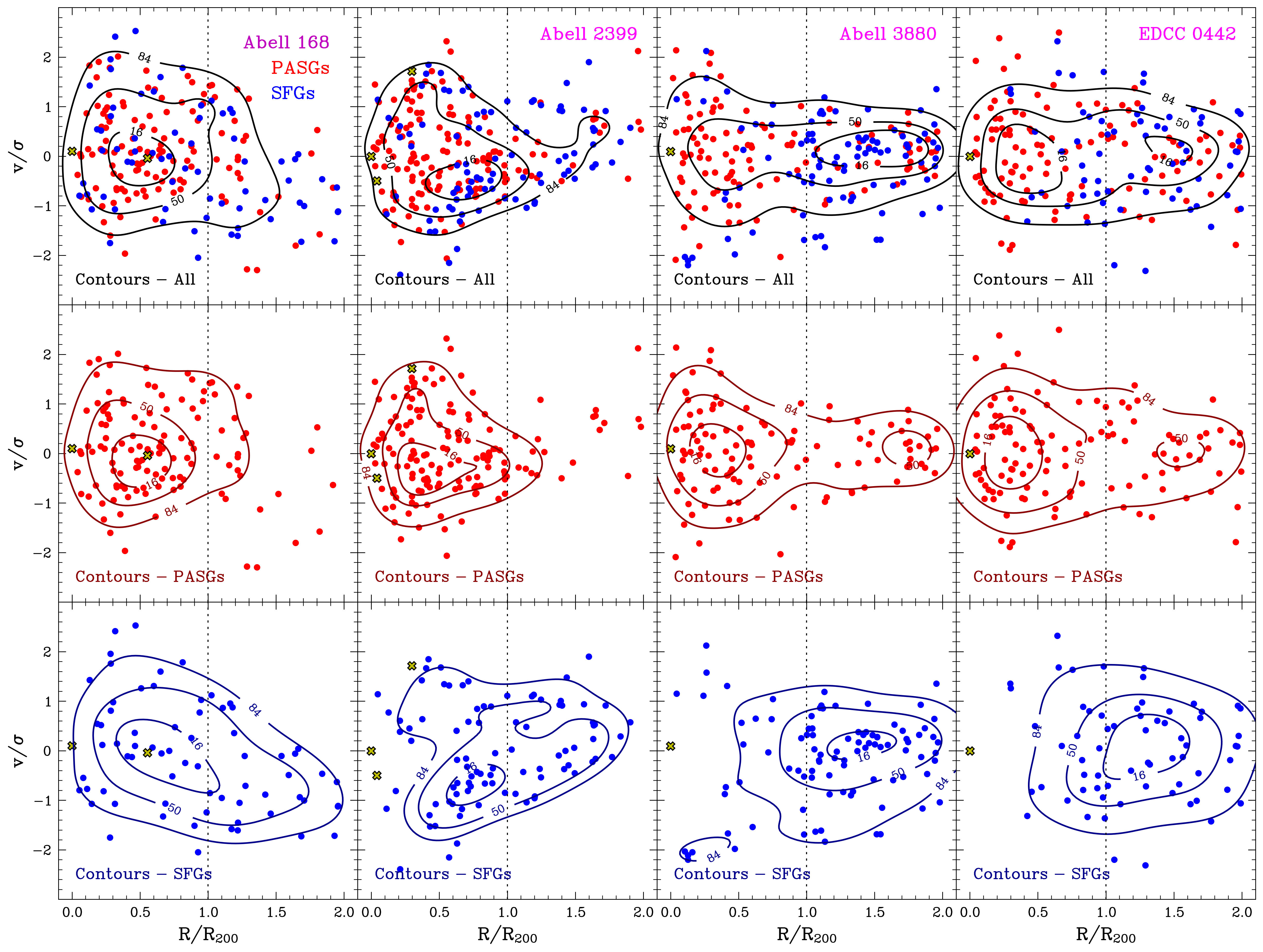}
    \caption{Same as Figure~\ref{Spatial_distribution}, but for projected phase-space distribution.}
    \label{Combined_PPS}
\end{figure*}
\vspace{2 mm}
\\
Figure~\ref{Combined_PPS} shows the projected phase-spaces for each cluster, highlighting each spectral class, where the red and blue points represent PASGs and SFGs, respectively. The positions of the BCGs are indicated by yellow crosses. As in Figure~\ref{Spatial_distribution}, each column represents an individual cluster, and each row shows a different population; the contours are also generated for 16$^{th}$, 50$^{th}$, and 84$^{th}$ percentiles with the \texttt{ks} package. 

Focusing on the top row of Figure~\ref{Combined_PPS}, which shows the PPS for both PASGs and SFGs, there is a clear difference in the PPS distribution of galaxies in clusters with different dynamical states. Considering the 50$^{th}$ percentile contour of the distribution, merging systems host galaxies mostly within \rtwo; relaxed systems have more extended galaxy distribution up to 2\rtwo - the distributions follow the trumpet shape of caustics observed in typical virialised systems \citep{Kaiser1987, Regos1989, Diaferio1997}. To test the significance of the differences across the sample, we use the multivariate two-sample KDE test introduced by \cite{Duong2012}. The KDE test is a non-parametric method in which the integrated mean square error is used to quantify the difference between two samples and has already been adopted to analyse PPS in the literature \citep{Lopez2017, deCarvalho2017, Owers2019, Costa2024}. The KDE test returns p-values of 0.311, 0.013, 0.193, 0.004, 0.089, and 0.485 for A168 vs A2399, A168 vs A3880, A168 vs EDCC 0442, A2399 vs A3880, A2399 vs EDCC 0442, and A3880 vs EDCC 0442, respectively, also summarised in Table~\ref{tab:p-values pps all}. The p-values are higher when clusters with similar dynamical states are compared, while they are lower when relaxed clusters are compared with merging clusters. However, A168 vs EDCC 0442 and A2399 vs EDCC 0442 result in p \gt 0.05. This is most likely due to a similar density peak (i.e., the contour of 16$^{th}$ percentile) observed in the region between \s0.3 \lt R/\rtwo \lt 0.7 and $\mid v/\sigma \mid$ \lt 0.5. For the stacked merger vs relaxed PPS comparison, the same analysis yields a lower p-value of 0.005.

\begin{table}
    \centering
    \begin{NiceTabular}{|c|c|c|c|}[corners=NW]
        \Hline
        \rowcolor{gray!0}
        & \Block{1-1}{A2399} & \Block{1-1}{A3880} & \Block{1-1}{EDCC 0442 }\\
        \Hline
        \rowcolors{gray!0}{}
        A168 & 0.311 & 0.013 & 0.193 \\
        \Hline
        A2399 & - & 0.004 & 0.089 \\
        \Hline
        A3880 & - & - & 0.485 \\
        \Hline
    \end{NiceTabular}
    \caption{p-values, returned by the KDE test, for cluster-to-cluster comparisons of the projected phase-spaces of all galaxies.}
    \label{tab:p-values pps all}
\end{table}

The PPS distribution for PASGs, shown in the middle row of Figure~\ref{Combined_PPS}, appears similar to the overall distribution shown in the top row. The PASGs mostly dominate smaller cluster-centric distances (and lower $\mid v/\sigma \mid$ values) consistent with the so-called "virialised" population \citep{Oman2013, Rhee2017}. However, the presence of this virialised region is more prominent for relaxed systems having a relatively symmetric velocity distribution around $v/\sigma \sim 0$ and a clear deficit of SFGs in the same region. Moreover, there are a few PASGs at R \gt \rtwo for merging systems, while relaxed systems contain a noticeable passive population in the outskirts seen as a secondary peak - as an indication of an infalling population. Cluster-to-cluster comparisons return p \gt0.05, indicating that PASGs exhibit similar distributions in the PPS. This also holds for the stacked samples, for which the KDE test returns p = 0.124.

\begin{figure*}[!t]
    \centering
    \includegraphics[width=\textwidth]{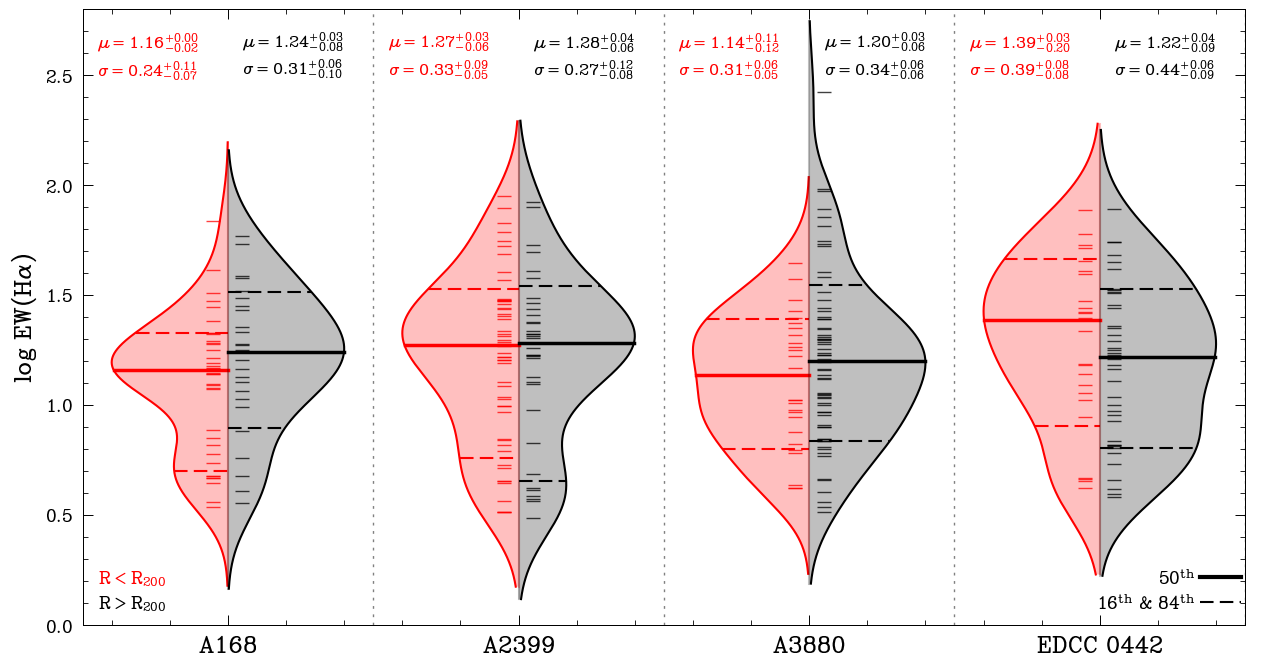}
    \caption{The EW(\ha) distributions. The red and black violin plots represent the distribution for R\lt \rtwo and 1\lt R/\rtwo \lt 2, respectively, for each cluster. On top of each violin distribution, the median and the standard deviation estimated via median absolute deviation (1.4826 $\times$ MAD) of EW(\ha) distribution are given and coloured to match the violin distributions. The uncertainties on these values are estimated through bootstrapping. The solid and dashed lines indicate the $50^{th}, 16^{th}$ and $84^{th}$ percentiles. The stripes represent the individual values for the given samples.}
    \label{EW_Halpha_histograms}
\end{figure*}

The PPS distribution for the SFGs shown in the bottom row of Figure~\ref{Combined_PPS} reveals intriguing differences between the merging and relaxed clusters. For the relaxed systems, the SFGs are found at larger cluster-centric distances, as shown in Section~\ref{Result - Radial distribution}, and have relatively symmetric distributions about the velocity axis. In contrast, the SFGs in the merging clusters are often found at small cluster-centric distances and show larger asymmetry in velocity. Comparisons between clusters reveal p-values from the KDE test of 0.089, 0.005, and 0.024 for A168 vs A2399, A168 vs A3880, and A168 vs EDCC 0442; 0.036, and 0.291 for A2399 vs A3880, and A2399 vs EDCC 0442; 0.374 for A3880 vs EDCC 0442, respectively (for R\lt2\rtwo), also summarised in Table~\ref{tab:p-values pps sfg}. These results highlight that clusters with similar merger states (i.e., A168 vs A2399 and A3880 vs EDCC 0442) demonstrate similarities in their PPS distribution among themselves. A168, as a post-merger cluster, presents an SFG phase-space distribution that is statistically different to A3880 and EDCC 0442. SFGs in A2399 show a similar distribution to EDCC 0442. However, their PPS distribution is inconsistent with the null hypothesis of being drawn from the same parent sample as A3880. In the case of analysis based on the stacked samples, the KDE test returns p = 0.014, inconsistent with the hypothesis that two samples are drawn from the same parent sample.

\begin{table}
    \centering
    \begin{NiceTabular}{|c|c|c|c|}[corners=NW]
        \Hline
        \rowcolor{gray!0}
        & \Block{1-1}{A2399} & \Block{1-1}{A3880} & \Block{1-1}{EDCC 0442 }\\
        \Hline
        \rowcolors{gray!0}{}
        A168 & 0.089 & 0.005 & 0.024 \\
        \Hline
        A2399 & - & 0.036 & 0.291 \\
        \Hline
        A3880 & - & - & 0.374 \\
        \Hline
    \end{NiceTabular}
    \caption{Same as Table~\ref{tab:p-values pps all}, but for SFGs.}
    \label{tab:p-values pps sfg}
\end{table}

\subsection{\texorpdfstring{EW(\ha) Distributions}{EW(Ha) Distributions}}\label{Result - EW(Ha) distribution}
In the previous sections, we showed that SFGs in mergers are more likely to be found in the region with ongoing dynamic activity. In this section, we investigate the mode of SF in order to understand if there is evidence for merger-related triggering or quenching of SF. To do this, we examine their distributions of EW(\ha) values, as tracers of ongoing SF, globally (Section~\ref{Global EW(Ha) Properties}) as well as by focusing on merger-specific regions (Section~\ref{Zoom in EW(Ha) Properties}).

\subsubsection{Global Comparison}\label{Global EW(Ha) Properties}
Figure~\ref{EW_Halpha_histograms} shows the distribution of log EW(\ha) for each cluster. Each red violin plot represents the distribution within \rtwo for a cluster, while the black violin plot shows the distribution for 1\lt R/\rtwo \lt 2. Above each plot, the median and the standard deviation determined from the median absolute deviation (1.4826 $\times$ MAD) of the EW(\ha) distribution are given and coloured to match the violin distributions. The uncertainties on these values are estimated through bootstrapping. The solid and dashed lines indicate the $50^{th}, 16^{th}$ and $84^{th}$ percentiles. The stripes represent the individual values for each sample. The median values representing each distribution do not differ significantly across the sample compared to the dispersion.
\\
\\
We apply the two-sample KS and AD tests to test for differences between the distributions of EW(\ha) within \rtwo and on the outskirts. For R\lt\rtwo, comparing the EW(\ha) distributions of the merging and relaxed clusters individually, we only find a statistically significant difference between A168 and EDCC 0442, where the p-values are 0.006 (KS) and 0.022 (AD), respectively. If we stack the relaxed clusters, this becomes less significant with $p_{KS} = 0.076$ and $p_{AD} = 0.142$. Comparisons of A168 versus A2399, and A3880 versus EDCC 0442 return p$_{KS/AD}$ = 0.032/0.054 and 0.081/0.044, indicating marginally significant differences between these samples. As we look at the outskirts (i.e., R \gt \rtwo), no statistically significant differences (i.e., $p\gg0.05$) are found across the sample. We also compare the distributions between galaxies within \rtwo and the outskirts per cluster and find no statistically significant differences ($p\gg0.05$) either.
\\
\\
The KS and AD tests also reveal no significant difference between the stacked merging and relaxed samples for R\lt\rtwo and R\gt\rtwo, returning $p\gg0.05$. \textit{These results indicate that there is no statistically significant difference in the global star-forming properties probed by EW(\ha) between the relaxed and merging clusters.}
\begin{figure*}[!t]
    \centering
    \includegraphics[width=\linewidth]{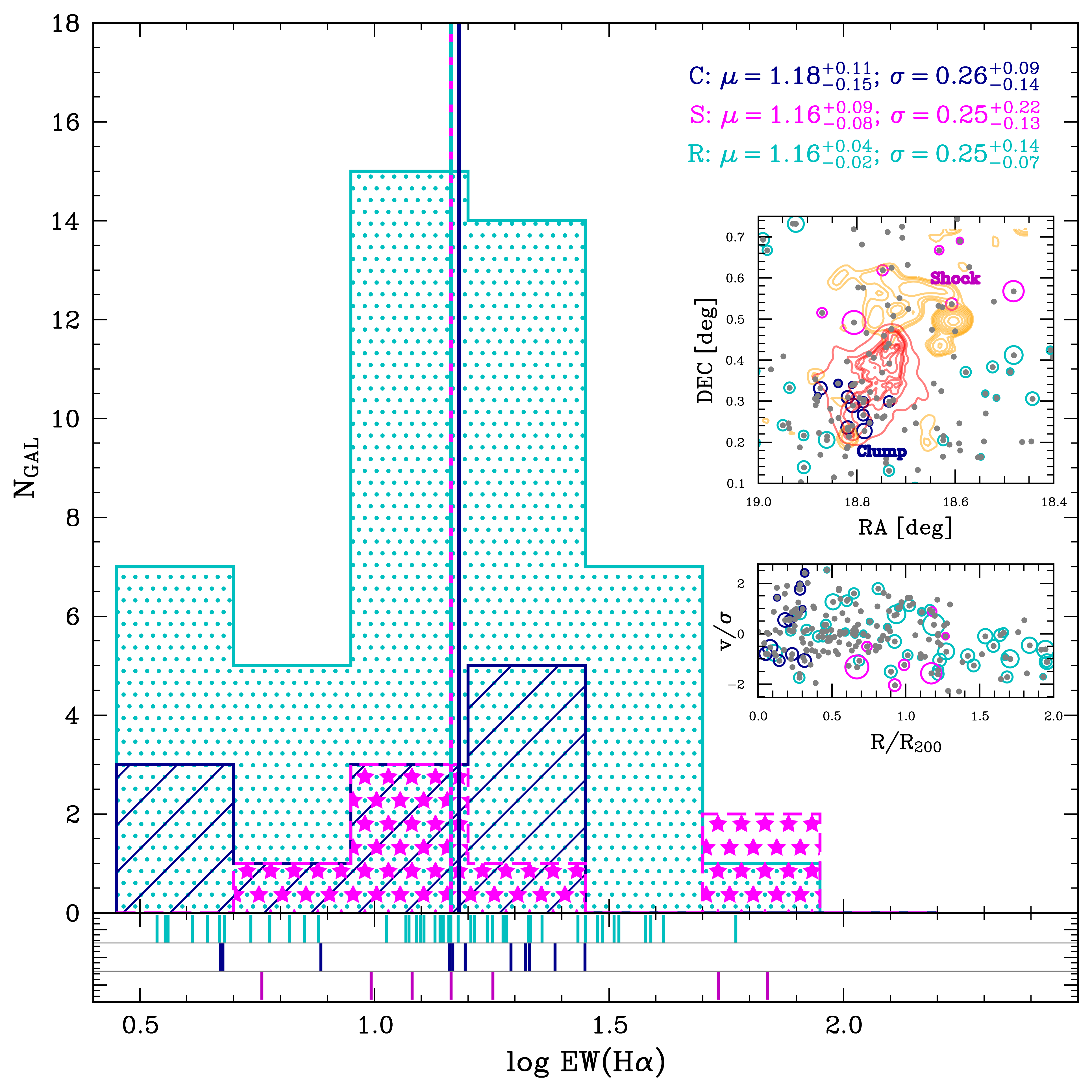}
    \caption{EW(\ha) distribution of subsamples defined in A168. The dark blue, magenta, and cyan symbols represent "Clump", "Shock", and "Rest" samples, respectively. The vertical lines mark the median of each subsample, which is given in the upper right of the figure with the standard deviation estimated via median absolute deviation (1.4826 $\times$ MAD) and uncertainties estimated via bootstrapping. The plot in the upper inset shows the spatial distribution of these samples close to the central region. The red and gold contours show the X-ray and radio emission from Figure~\ref{Xray_maps}. The symbol sizes are proportional to the EW(\ha); the larger the symbol, the higher the value. The lower inset shows the projected phase-space distributions. The stripes shown in the bottom panel represent the individual values for each sample.}
    \label{EW_Ha_merger_regions_A168}
\end{figure*}

\subsubsection{A closer look into merger regions}\label{Zoom in EW(Ha) Properties}
As shown in Section~\ref{Global EW(Ha) Properties}, there is no clear evidence for global enhancement or quenching of SF in the merging clusters. To detect any possible merger-induced signal, we now focus on the galaxies situated in the vicinity of merger regions (e.g., between substructures) and features such as radio relics (i.e., shocks) to compare them with the remaining SFGs. We define subsamples of SFGs based on their spatial and/or PPS distribution. 

\begin{figure*}[!t]
    \centering
    \includegraphics[width=\linewidth]{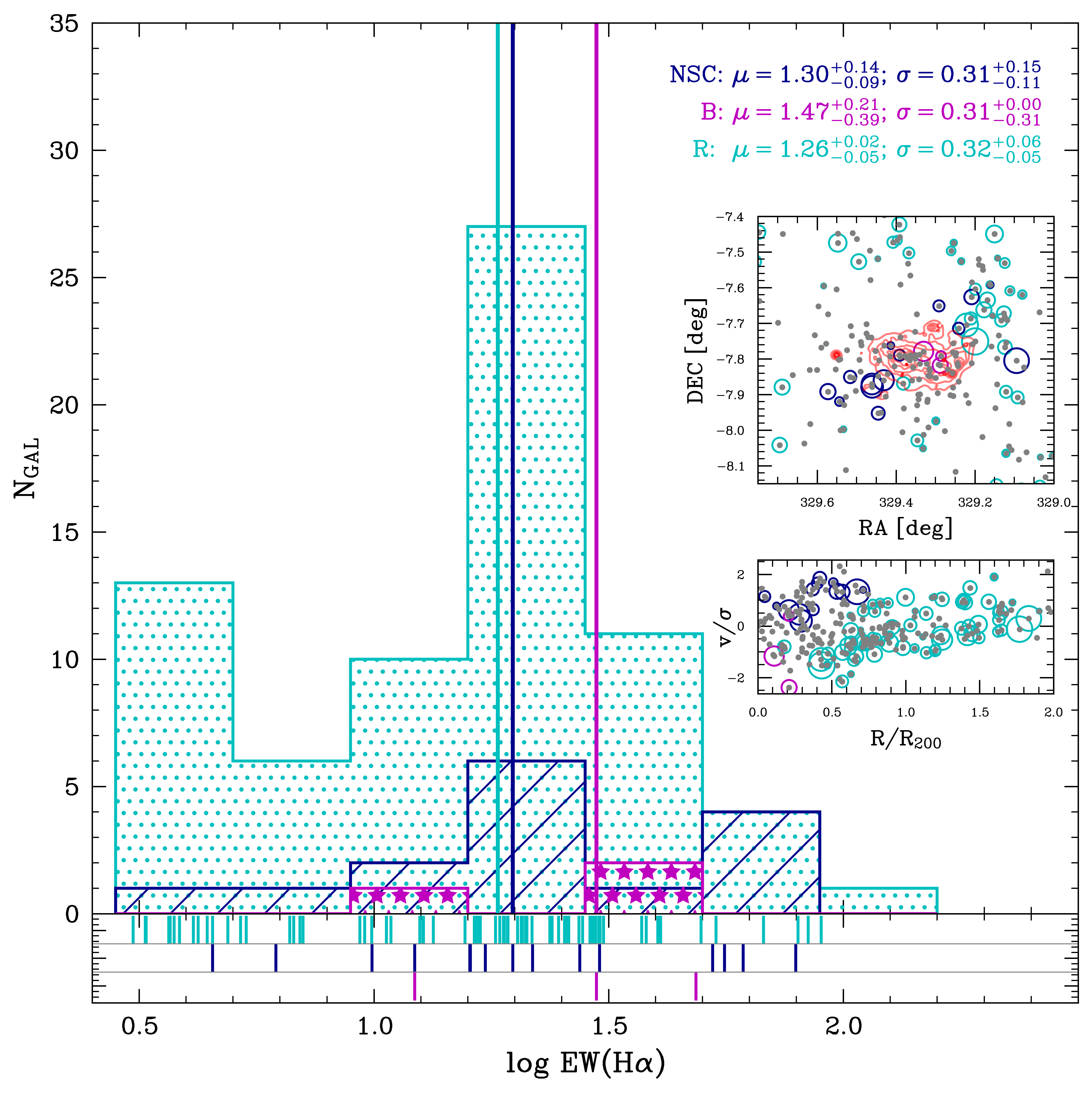}
    \caption{Same as Figure~\ref{EW_Ha_merger_regions_A168}, but for A2399. The dark blue, magenta, and cyan symbols represent "NSC", "B", and "Rest" samples, respectively. The vertical lines are the medians of each subsample and given in the upper right of the figure with the standard deviation estimated via median absolute deviation (1.4826 $\times$ MAD). The inset in the upper right highlights the central region, showing the spatial distribution of the samples in greater detail. The red contours are for the X-ray emission shown in Figure~\ref{Xray_maps}.}
    \label{EW_Ha_merger_regions_A2399}
\end{figure*}

\begin{itemize}[leftmargin=0.1in]
  \item \textbf{A168:}
  \begin{itemize}
    \item \textbf{"Clump" galaxies (C)}: Galaxies located between substructures (N=12).
    \item \textbf{"Shock" galaxies (S)}: Galaxies identified near the shock/relic (N=7).
    \item \textbf{"Rest" galaxies (R)}: All remaining galaxies not in "C" or "S" (N=50).
  \end{itemize}
  
  \item \textbf{A2399:}
  \begin{itemize}
    \item \textbf{"Near-subcluster" galaxies (NSC)}: Galaxies close to the western clump in the PPS distribution (N=15).
    \item \textbf{"Between XP" galaxies (B)}: Galaxies located between substructures (i.e., between X-ray peaks) (N=3).
    \item \textbf{"Rest" galaxies (R)}: All remaining galaxies not in "NSC" or "B" (N=72).
    \item One galaxy (ID: 9239900059) is common to both "B" and "NSC" categories.
  \end{itemize}
\end{itemize}

Figure~\ref{EW_Ha_merger_regions_A168} shows the log-scaled EW(\ha) distribution of SFGs in A168. The dark blue, magenta, and cyan histograms represent the galaxies in the clump, shock, and rest samples. The inset plot on the upper right highlights the spatial distribution of these populations by zooming into the central region; the red and orange contours display the X-ray and radio maps; the inset plot on the middle right shows their distribution in the PPS; the bottom panel displays the stripe density plot for each population with the same colour code. In the upper right corner, median and standard deviation determined from MAD are given per subsample, and the associated uncertainties are estimated in a similar fashion to that described previously. The median values are nearly identical, and the dispersions are similar (the conventional mean returns 1.10 dex, 1.26 dex, and 1.15 dex for the clump, the shock, and the rest). However, despite having a smaller sample size, shock galaxies extend toward the higher end of the distribution; \s30\% (2/7) are in the top 5\% of EW(\ha) values of \s69 \AAs(the highest), and \s54 \AAs(third highest). The probability of 2 or more of the 7 lying in the 5\% tail is 0.044. The galaxy with the highest EW(\ha) is located at the SE end of the shock, which suggests SF may be triggered by being overrun by the shock.
\\
\\
To determine the significance of the difference between subsamples, we run internal and external comparisons. The clump and shock galaxies are first compared with the remaining SFGs (after excluding each subsample, respectively) in A168 for R\lt\rtwo and R\gt\rtwo and then with each other. For external comparisons, we select SFGs within the same cluster-centric regions for A2399 and the stacked relaxed sample. Through the KS and AD tests, we find no significant difference for any comparison, i.e., all $p\gg0.05$.
\\
\\
Figure~\ref{EW_Ha_merger_regions_A2399} shows the EW(\ha) distributions of the subsamples defined in A2399. The dark blue, magenta, and cyan histograms, vertical lines, and circles represent the "NSC", "B", and the "rest" samples, respectively. The contours shown in the inset plot display the X-ray data. The medians and standard deviation values estimated from MAD are shown at the upper right for each subsample in the same colour and the uncertainties are estimated via bootstrapping. The samples of "NSC" and "B" show slightly higher values (1.30 dex and 1.47 dex) than the rest (1.26 dex). Yet, like A168, allowing for the dispersions, we find no significant differences between subsamples.
\\
\\
We conduct a similar set of comparisons for the subsamples defined in A2399. We only detect a significant difference between A2399's NSC subsample and SFGs residing within R\lt\rtwo in A168 with p-values of 0.049 and 0.038 from the KS and AD tests, respectively, which might hint at a difference between their pre-and post-merger nature.
\\
\\
Lastly, we follow up on the significant difference found for A168 and A2399 by comparing the subsamples defined in both clusters. There is no significant difference revealed by either statistical test. Only the AD test yields a marginal difference with $p = 0.071$ for the comparison between A168's "C" and A2399's "B" subsample.
\\
\\
Overall, our analysis of the distributions of EW(\ha) in the SFGs provides no evidence for SF triggered by the merger in the central regions of these clusters.

\section{Discussion} \label{sec:Discussion}
The main goal of this study is to understand the interplay between the SF activity of cluster galaxies and their hosts' assembly histories, in order to disentangle any additional impacts to those seen in the virialised/relaxed cluster environments. SF activity was traced through the $f_{SFG}$ and the distributions of EW(\ha). While SFGs in merging clusters exhibit spatially and kinematically more mixed distributions, the EW(\ha) properties are similar to those of relaxed systems, leading us to a scenario of dynamical mixing rather than enhancement in SF activity. In this section, we discuss our interpretation in the context of the existing literature and from a simulation point of view, as well as the limitations of the method and data.

\subsection{Comparison with the previous studies}\label{subsec:Discussion - Literature}
Combining the results presented in Section~\ref{Result - Fractions}, Section~\ref{Result - Spatial distribution}, and Section~\ref{Result - Radial distribution}, we are able to reproduce the well-established radial trend where the cluster centres are mainly dominated by PASGs, while SFGs are more common in the outskirts, agreeing with the previous studies \citep{Dressler1980, Biviano1997, Lewis2002, vdL2010, Barsanti2018, Owers2019}. This implies that the trend is not significantly affected by the dynamical states of the clusters. A similar result is reported by \cite{Astudillo2025} that the morphology/SFR-density relation still holds for unrelaxed clusters.
\\
\\
The mild increase found in f$_{SFG}$ (\s20\%) for merging clusters is comparable with findings in \citet{Cohen2014, Cohen2015} and \citet{Yoon2020}. They studied more than 100 nearby ($z<0.2$) SDSS clusters and found that clusters having substructures exhibit enhancement in the $f_{SFG}$ (\s20-30\%). However, our signal is less prominent ($\sim1.2\sigma$) compared to theirs ($\gtrsim2\sigma$), likely due to the difference in sample sizes. \cite{Stroe2017} studied 19 intermediate redshift clusters and 3000 associated \ha emitters, and the merging clusters, traced by radio shocks, were found to host more \ha emitters relative to relaxed clusters. More recently, \cite{Aldas2024} also reached similar conclusions based on \textsc{\texttt{Illustris TNG100}} simulation. Accordingly, our result on $f_{SFG}$ is qualitatively in agreement with the earlier works. Furthermore, the cluster outskirts present similar $f_{SFG}$, regardless of cluster dynamics, indicating that the outskirts are more resilient to dynamical state. This contradicts \cite{Aldas2024}, who found the majority of SFGs in the disturbed clusters reside in the outskirts (R\gt\rtwo).
\\
\\
Furthermore, we could not find any evidence of a recently quenched population as an indication of ongoing quenching. HDSGs make up a very small fraction of our sample, being represented by only 3 galaxies, which corresponds to \s0.36\% of the total sample (3/829). This is significantly lower than that found in WINGS clusters by \cite{Paccagnella2017} (i.e., \s7.3\%). \cite{Owers2019}, exploiting the spatially resolved data from SAMI-GS, found 17 HDSGs for the SAMI cluster sample, considering galaxies only with $\log(M_\ast/M_\odot)>10$ within \rtwo (N=579 galaxies), resulting in a fraction of \s3\%. They estimated that no more than \s0.5\% would be recovered as \hd-strong in single-fibre spectroscopy, agreeing with our HDSG fraction. Moreover, EDCC 0442, as a relaxed system, is the only cluster exhibiting a clear absence of SFGs within 0.5\rtwo. Compared to a strong deficit in SFGs found in post-merger A520 by \cite{Deshev2017}, we are not able to see any evidence for elevated quenching in merging systems. 
\\
\\
To search for any merger-induced SF activity, we examined the EW(\ha) distributions for both global cluster environments and in the vicinity of merger features. Even though the global EW(\ha) properties show marginal cluster-to-cluster differences, no evident trend as a function of dynamical state is found due to the large scatter in the distributions. Figures 8 and 9 of \cite{Stroe2017} show that, despite having a higher characteristic \ha density, merging clusters display similar \ha luminosities (proportional to SFR or EW(\ha) in our case) to relaxed clusters, in agreement with our result on global EW(\ha) properties. Moreover, \cite{Aldas2024} show that the sSFR distribution within \rtwo is similar for relaxed and disturbed clusters, also complementing our result. Close-up investigation of the galaxies near the merger feature does not indicate any strong merger-induced activity for our merging clusters. For A168, \cite{HL2009} investigated the angular variance of galaxies between substructures (i.e., the "clump" sample), and projection effects are unlikely to account for this clustering of SFGs, suggesting it is due to merger-induced activity. However, our EW(\ha) comparisons involving the "clump" galaxies indicate no significant difference compared to the remaining SFGs (including the relaxed systems), so that there is no strong evidence for merger-related triggering of SF. On the other hand, galaxies near the shock in A168 present a wider range of EW(\ha). Two of them exhibit high EW(\ha), a hint of merger-induced activity, consistent with \citet{Owers2012} and \cite{Stroe2015, Stroe2017}. Due to the low number statistics, this is not conclusive evidence for induced SF activity.
\\
\\
Regarding the origin of the mild enhancement in $f_{SFG}$ in merging clusters across our sample, we suggest a dynamical mixing scenario (i.e., redistribution of galaxies) where merging clusters represent an intermediate stage of cluster formation, being fed by star-forming rich filaments/groups. This explanation is also preferred by \cite{Cohen2015}. Figure 10 of \cite{Owers2017} illustrates the probable substructures in A2399 in addition to the western cluster. The peak in SFGs shown in Section~\ref{Result - Spatial distribution} and the SE extension of the "NSC" subsample coincide with these substructures, indicating multiple ongoing coalescences. Given the fact that A2399 is a dynamically young pre-merger system sitting at the centre of the Aquarius-Cetus Supercluster \citep{Bregman2004}, it is expected to see the infall of SF-rich galaxies or groups. Similar results are found by \citet{Cortese2004} and \citet{Deshev2017} for the merging systems A1367 and A520, respectively, where infalling substructures exhibit a higher $f_{SFG}$. An earlier study by \citet{Fujita1999a} using numerical simulations showed that due to the high relative velocities, blue SFGs can reach the core region of the merging clusters before they quench, supporting the presence of SFGs in the cores of merging clusters.

\subsection{Simulation perspective}\label{subsec:Discussion - Simulation}
\begin{figure*}[!t]
    \centering
    \includegraphics[width=\textwidth]{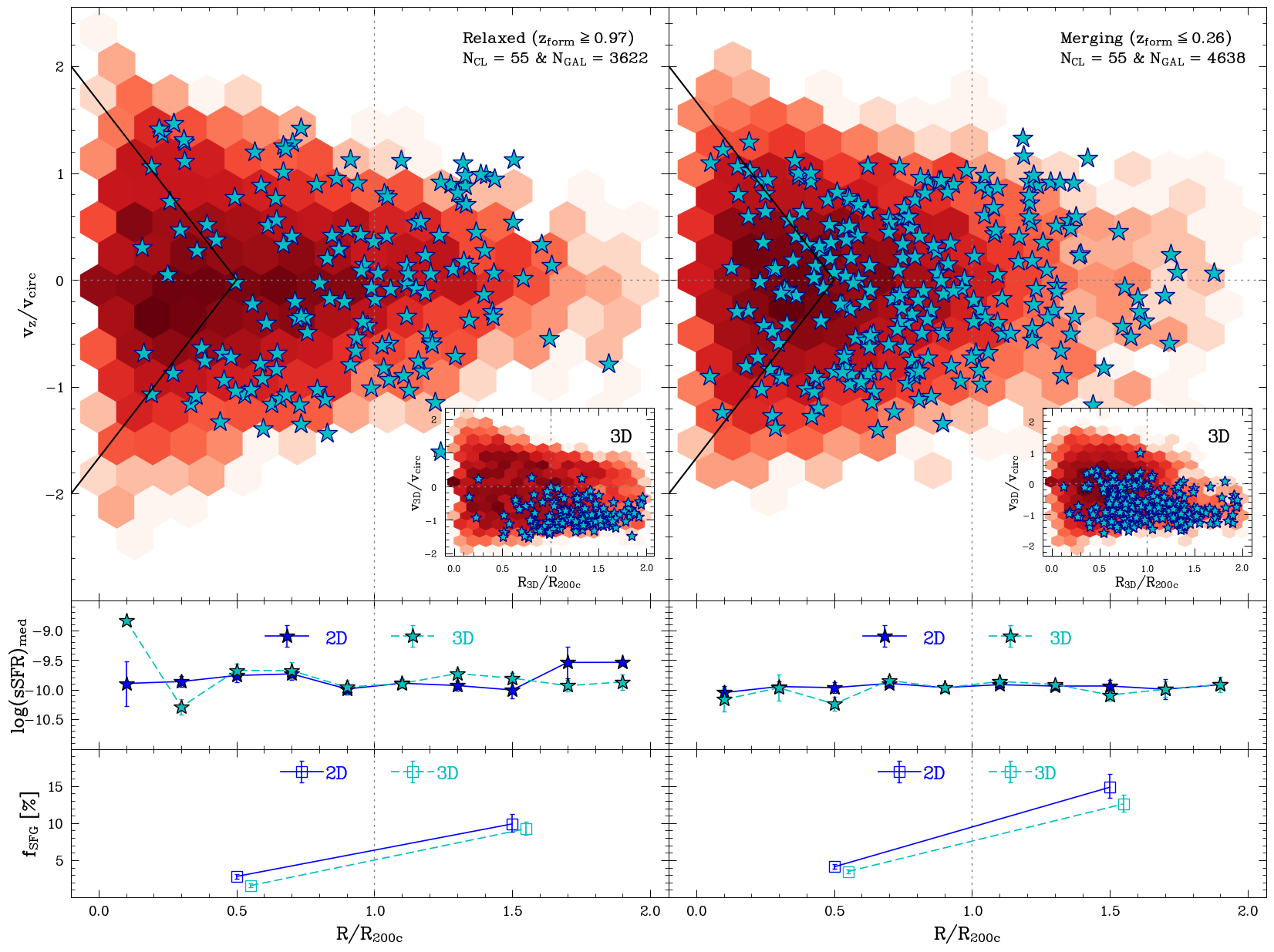}
    \caption{Magneticum simulation results for relaxed (left) and merging (right) clusters. \textbf{Top panel:} The xy projected phase-space distributions. The inset plots show the 6D phase-space diagram using the radial component of the 3D velocity and the 3D radius. The red hexagonal bins show the passive galaxies; the redder the colour higher the density; the cyan stars represent star-forming galaxies for each plot. The black solid lines define the virialised regions \citep{Rhee2017} for each panel. \textbf{Middle panel:} The median sSFR as a function of normalised cluster-centric distance. The blue solid line with stars represents the sSFR profile based on the xy projection (2D), while the cyan dashed line with stars is the same for the real (3D) space. The uncertainties are standard errors of medians (i.e., $1.253 \times \sigma_{bin} / \sqrt{N_{bin}}$). Note that the central bin of the 3D sSFR profile for relaxed clusters (left panel) contains only 1 galaxy. \textbf{Bottom panel:} The star-forming galaxy fractions as a function of normalised cluster-centric distance. While the colours are the same as in the middle panel, we use open squares for the symbols to improve visual clarity. The error bars show $1\sigma$ uncertainties estimated through the recipe given in \citet{Cameron2011}. For each panel, the grey vertical dotted line indicates \rtwo. The horizontal grey dotted lines shown in the top panels are centred at 0.}
    \label{Simulation_results}
\end{figure*}
To validate our interpretation for the enhancement in the incidence of SFGs in merging clusters from a simulation perspective, we utilise the Box2-hr volume from the Magneticum Pathfinder\footnote{\url{www.magneticum.org}} suite of hydrodynamical cosmological simulations. The simulations were run with a modified version of \textsc{GADGET-2} \citep{springel05} with various improvements \citep{dolag05,beck16}. The large volume for Box2~hr, of $(352\mathrm{Mpc}/h)^3$, includes a significant number of galaxy clusters to compare with. For more details on the simulations, including particle resolution, structure identification, baryonic physics, galaxy cluster populations, and cluster galaxy populations, see \cite{spingel01,dolag09,teklu15, remus17, lotz21}.
\\
\\
We chose snapshot~$140$ at $z=0.03$, comparable to our observed sample, where there are $886$~clusters with halo masses of $M_\mathrm{200c}\geq 10^{14} M_\odot$, reaching as high as $M_\mathrm{200c}\approx 2\times 10^{15} M_\odot$. This sample is reduced to the $868$~clusters with unbroken assembly histories up to $z=3$, equivalent to the sample used by \citet{Kimmig2025} to study the link between the distribution of stellar mass in galaxy clusters and their formation histories. We consider the galaxy populations of these clusters with a stellar mass cut of $M_*\geq 5\times10^{9}M_\odot$, for a total number of 45902 galaxies in the clusters. In order to cover the same mass regime, we also apply a cut on cluster mass, excluding those outside of the SAMI range (i.e., $14.25\lesssim \log M_{200}\lesssim15.19$), leaving 340 clusters. 
\\
\\
The large number provides a huge range of assembly histories, with formation redshifts (defined as the time when the main progenitor of the cluster first reached $50$\%~of the final $z=0$ halo mass) between $z_\mathrm{form}=0.058$ and $z_\mathrm{form}=1.77$, with a median formation at $z_\mathrm{form}\approx0.67$. We split the sample into early and late forming by the $1\sigma$-bounds of the $z_\mathrm{form}$ distribution of the remaining 340 clusters. The cuts are therefore $z_\mathrm{form}\lesssim 0.26$ ($t_\mathrm{lookback}\simeq3\ Gyr$) for dynamically active clusters and $z_\mathrm{form}\gtrsim0.97$ ($t_\mathrm{lookback}\simeq8\ Gyr$) for dynamically relaxed clusters. $z_\mathrm{form}<0.26$ is well-representative of cases where a system experiences a merger within the last \s3 Gyr (which is quite comparable with A168). After applying the cuts, the size of the final sample size is reduced to 55 merging and 55 relaxed clusters with 4638 and 3622 members, respectively.
\\
\\
Figure~\ref{Simulation_results} shows the results for the Magneticum clusters. The top panel shows the stacked projected phase-space distributions (in the 2D x-y projection; normalised $R_{xy}$ and $v_{z}$) of each cluster sample, whereas inset plots are based on the galaxies' 6D properties (i.e., real distance, $R_{3D}$; velocities, $v_{3D}$, with respect to the centre). The hexagonal bins represent PASGs; the redder the colour, the higher the phase-space density per bin, and SFGs are plotted as cyan stars. Here, $v_{circ}$ denotes the circular velocity ($\sqrt{GM/r}$) of the halo as a tracer of velocity dispersion, and $R_{200c}$ is the radius within which the average mass density is 200 times the critical mass density at the given redshift. In the middle panel, the blue stars with a solid line and the cyan stars with a dashed line display the median sSFR as a function of projected and 3D distances, respectively. The uncertainties are the standard errors on the median (i.e., $1.253\times\frac{\sigma_{bin}}{\sqrt{N_{bin}}}$). The cyan and blue open squares in the bottom panel highlight projected and 3D $f_{SFG}$ within R\lt\rtwo and 1 \lt R/\rtwo \lt 2 for each sample, respectively, and are shifted radially in order to avoid overlapping. The error bars in the bottom row indicate 1$\sigma$ uncertainties estimated through \citet{Cameron2011}. The grey vertical dashed lines denote \rtwo for all panels. The horizontal dashed lines are centred at $v/v_{circ} = 0$ for the top panels. The black solid lines in the top row of Figure~\ref{Simulation_results} mark the region defined by \citet{Rhee2017}, primarily enclosing the ancient infaller or so-called virialised population. As merging cluster cores may not be fully virialised, this region may not represent the true virialised population. \citet{Rhee2017} caution that, in equal-mass mergers, adopting a single cluster centre can introduce PPS scatter. With these caveats, we retain the term "virialised" to refer to the central/core regions of these systems.
\\
\\
The simulated PPSs show striking similarities to our observations. The relaxed clusters host a few SFGs in the virialised region, whereas the presence of SFGs is more pronounced in the same space for merging clusters. In order to rule out possible projection effects, we also show 6D phase spaces. SFGs in relaxed clusters are mainly consistent with an infalling population \citep[mostly first infallers;][]{Rhee2017}, where the majority of SFGs contribute to this population. This aligns with \citet{Lotz2019}, who found that the majority of SFGs do not survive beyond the first core passage. In contrast, the merging systems host nearly half of their SFGs in the inner regions represented by intermediate and ancient infallers \citep{Rhee2017}. Regardless of the projection, these results support that the merging clusters host a more mixed SFG population and exhibit slightly higher $f_{SFG}$ within \rtwo (bottom panel of Figure~\ref{Simulation_results}) compared to the relaxed systems. 
\\
\\
Consistent with our findings, the simulated merging clusters also seem to have a high incidence of SFGs within \rtwo, where the impact of the merger is greatest. In order to test whether this is because of possible merger-induced SF activity or not, we now investigate the SF activity of SFGs in both relaxed and merging clusters. We trace SF activity via median sSFR values as a function of cluster-centric distance with a step size of $0.2R/R_{200c}$ for the simulated clusters. Both samples exhibit a similar trend in median sSFR with a relatively constant radial profile for projected and 3D spaces. The two innermost bins for 3D space for the relaxed clusters are represented by just three galaxies, so are not reliable. Combined with the PPS properties, we can conclude that the constant radial trend in sSFR indicates that cluster mergers have little impact on global SF activity. In other words, there is no clear evidence for triggered SF activity, supporting our interpretation that the differences are due to dynamical mixing rather than enhanced activity. Moreover, given the fact that relaxed clusters, representing the dynamically more evolved, passive, older systems (i.e., higher $z_\mathrm{form}$), have been suppressing the SF over longer timescales compared to young merging systems, this may also reflect as the difference in $f_{SFG}$. Recently, \citet{Kimmig2025} found a positive correlation (shown in their Figure 9), although a weaker relation, between quenched galaxy fractions and formation redshift for different simulations, including Magneticum, which also supports this.

\subsection{Caveats and Limitations}
In cluster environments, RPS mainly regulates SF activity in an outside-in manner as the galactic gravitational potential weakens toward the outskirts \citep{Cortese2021}. It presents observationally by forming stripped star-forming tails, recently quenched regions, and truncated disks \citep{Koopmann2004, Boselli2006b, Chung2009, Smith2010, Poggianti2016, Owers2019, Bellhouse2022}. This means that any enhancement/quenching in SF due to RPS is likely to appear in the outskirts of galaxies. The results presented in this study rely on single-fibre spectroscopy from SAMI-CRS using the 2dF instrument (2 arcseconds fibre diameter) and SDSS (3 arcseconds). Considering the redshift range (0.04 \lt z \lt 0.06), the coverage of the spectra only ranges over physical scales of \s1.6 and 2.4 kpc. This means that we are only able to probe the very central regions of galaxies and can not retrieve information from the outskirts, where any signal due to environmental effects might be found. In this regard, the resolved spectroscopy from the SAMI and  Hector Galaxy Surveys will allow SF activity to be investigated over larger regions.
\\
\\
The sample studied here only consists of four clusters. Studies based on a single or handful of merging systems present contradictory results \citep{Ferrari2005, HL2009, Owers2012, Pranger2014, Deshev2017}, indicating that cluster-to-cluster variations affect the overall interpretation. However, a handful of studies based on larger datasets have provided a hint of merging clusters triggering SF or hosting more SFGs \citep{Cohen2014, Cohen2015, Yoon2020, Lourenco2023}. Therefore, disentangling the impact of merging clusters requires more complete, statistically robust studies over larger cluster samples, with a well-covered range of dynamical states.

\section{Summary and Conclusions}\label{sec:Conclusion}
Many questions concerning galaxy evolution in extreme merging clusters remain open. This paper addresses some of these using a sample of four nearby SAMI clusters ($0.04 < z <0.06$) with different dynamical activity - A168, A2399, A3880, and EDCC 0442. Utilising spectra from the SAMI Cluster Redshift Survey and Sloan Digital Sky Survey, galaxy populations were classified as passive, star-forming, and \hd-strong based on spectral features, through a classification scheme outlined in Section~\ref{sec:Spectral classification}. We have investigated their spatial and projected phase-space distributions, and the EW(\ha) properties, as a tracer of SF activity, with a focus on SFGs. Our key findings are the following:

\begin{enumerate}
    \item We found a mild increase (\s20\%) in $f_{SFG}$ within \rtwo for the merging clusters compared to their relaxed counterparts in our sample.  Yet, the difference is within a 1$\sigma$ level.   

    \item Spatial distributions and projected phase-space diagrams reveal differences between merging and relaxed clusters. In the relaxed clusters, there is clear segregation between the PASG and SFG populations, while the galaxy populations are better mixed in the merging clusters, with the majority of SFGs found within \rtwo. 
    
    \item Despite the mixing, the well-defined relation between SF activity and cluster-centric distance (i.e., SF-density) is still visible for the merging clusters.
    
    \item We found no significant difference in the properties of EW(\ha) between merging and relaxed clusters, both globally and locally, ruling out the merger-induced activity.

    \item Comparison with the Magneticum simulations indicates that the additional SFGs in the central regions of merging clusters are most likely undergoing their first infall into the merged system.
\end{enumerate}

We conclude that the mild enhancement seen in the $f_{SFG}$ in the merging clusters is probably due to the redistribution (i.e., mixing) of galaxies in dynamically young environments that are in an intermediate stage of formation; whereas the relaxed counterparts, which represent a dynamically older population of clusters, have had much more time to suppress SF in their members.
\\
\\
The contradictory results reported in the literature emphasise the need to study a large sample of clusters with well-characterised dynamical activity. To that end, upcoming surveys and missions such as 4MOST Hemisphere Survey \citep[\texttt{4HS};][]{4HS} and \texttt{NewAthena} \citep{Athena} will provide multi-wavelength data across different environments and enable us to study statistically the impact of varying dynamical states on galaxy properties with a larger sample. 
\\
\\
In the future, we will investigate the environmental impact on SF activity using resolved data from SAMI-GS, which will enable us to search for direct evidence for triggering/quenching in SF (Çakır et al., submitted). Moreover, the next-generation Hector Galaxy Survey \citep{Bryant2024, Oh2025} will provide resolved data for a larger cluster sample, including more massive systems with more diverse dynamical activity, out to 2\rtwo, where we will be able to address the same questions with spatially resolved galaxy properties. 

\begin{acknowledgement}
We thank the anonymous referee for their constructive feedback that has helped to improve the clarity of the manuscript. We thank Luca Cortese for his valuable feedback on this manuscript. We also thank Stefania Barsanti for her feedback on the discussion section. 

OÇ and GQ are supported by the Australian Government Research Training Program Scholarship (RTP) to conduct this research. LCK acknowledges support by the Deutsche Forschungsgemeinschaft (DFG, German Research Foundation) under project nr. 516355818, the COMPLEX project from the European Research Council (ERC) under the European Union’s Horizon 2020 research and innovation program grant agreement ERC-2019-AdG 882679 and by DFG under Germany's Excellence Strategy -- EXC-2096 -- 3900783311. This research was partially supported by the Australian Research Council Centre of Excellence for All Sky Astrophysics in 3 Dimensions (ASTRO 3D), through project number CE170100013. This work was supported by the Korea Astronomy and Space Science Institute under the R\&D program (Project No. 2025-1-831-01) supervised by the Ministry of Science and ICT (MSIT). MP acknowledges support from the National Research Foundation of Korea (NRF) grant funded by the Korea government (MSIT) (No. 2022R1A2C1004025).

The SAMI-CRS made extensive use of the Anglo-Australian Telescope (AAT). We acknowledge the traditional custodians of the land on which the AAT stands, the Gamilaraay people, and pay our respects to elders past and present, and all the technical and observing support of the staff at the AAT throughout the SAMI-CRS.

Funding for SDSS-III has been provided by the Alfred P. Sloan Foundation, the Participating Institutions, the National Science Foundation, and the U.S. Department of Energy Office of Science. The SDSS-III web site is http://www.sdss3.org/.

SDSS-III is managed by the Astrophysical Research Consortium for the Participating Institutions of the SDSS-III Collaboration including the University of Arizona, the Brazilian Participation Group, Brookhaven National Laboratory, Carnegie Mellon University, University of Florida, the French Participation Group, the German Participation Group, Harvard University, the Instituto de Astrofisica de Canarias, the Michigan State/Notre Dame/JINA Participation Group, Johns Hopkins University, Lawrence Berkeley National Laboratory, Max Planck Institute for Astrophysics, Max Planck Institute for Extraterrestrial Physics, New Mexico State University, New York University, Ohio State University, Pennsylvania State University, University of Portsmouth, Princeton University, the Spanish Participation Group, University of Tokyo, University of Utah, Vanderbilt University, University of Virginia, University of Washington, and Yale University.

This study made extensive use of \texttt{Python} \citep{Python3} libraries --- \textsc{\texttt{Astropy}} \citep{Astropy2013, Astropy2018, Astropy2022}, \textsc{\texttt{Numpy}} \citep{Numpy}, \textsc{\texttt{Scipy}} \citep{Scipy},
\textsc{\texttt{statsmodels}} \citep{statsmodels},
\textsc{\texttt{Matplotlib}} \citep{Matplotlib}, \textsc{\texttt{Smplotlib}} \citep{Smplotlib}, \textsc{\texttt{Pandas}} \citep{Pandas}, \textsc{\texttt{Seaborn}} \citep{Seaborn}, and \textsc{\texttt{Jupyter}} \citep{Jupyter}; and \textsc{\texttt{R}} \citep{R}.

\end{acknowledgement}

\printendnotes

\bibliography{Bibliography}

\end{document}